\documentstyle[11pt,leqno,bezier]{article}

\newtheorem{theorem}{Theorem}[section]
\newtheorem{definition}[theorem]{Definition}
\newtheorem{corollary}[theorem]{Corollary}

\newtheorem{example}[theorem]{Example}

\newtheorem{proposition}[theorem]{Proposition}

\long\def\comment#1\endcomment{{}}

\comment Martin's Personal Macros \endcomment

\catcode`\@=11
\catcode`\+=14
\def\@begintheorem#1#2{\it \trivlist \item[\hskip
 \labelsep{\bf #1\ #2.}]}
\def\@opargbegintheorem#1#2#3{\it \trivlist\item[\hskip+
 \labelsep{\bf #1\ #2.\ (#3)}]}
\def\@endtheorem{\endtrivlist}

\def\@listI{\leftmargin\leftmargini \parsep 1pt plus 2.5pt
 minus 1pt\topsep 10pt plus 4pt minus 6pt+
 \itemsep 0pt plus 2.5pt minus 1pt}
\let\@listi\@listI
\@listi

\def\@sect#1#2#3#4#5#6[#7]#8{\ifnum #2>\c@secnumdepth+
 \def \@svsec {}\else \refstepcounter {#1}\edef \@svsec+
 {\csname the#1\endcsname. \hskip .1em }\fi \@tempskipa+
 #5\relax \ifdim \@tempskipa >\z@ \begingroup #6\relax+
 \@hangfrom {\hskip #3\relax \@svsec }{\interlinepenalty+
 \@M #8.\par }\endgroup \csname #1mark\endcsname {#7}+
 \addcontentsline {toc}{#1}{\ifnum #2>\c@secnumdepth+
 \else \protect \numberline {\csname the#1\endcsname. }+
 \fi #7}\else \def \@svsechd {#6\hskip #3\@svsec #8.+
 \csname #1mark\endcsname {#7}\addcontentsline {toc}{#1}+
 {\ifnum #2>\c@secnumdepth \else \protect \numberline+
 {\csname the#1\endcsname. }\fi #7}}\fi \@xsect {#5}}

\def\section{\@startsection {section}{1}{\z@ }+
 {-3.5ex plus -1ex minus -.2ex}{2.3ex plus .2ex}{\bf }}
\def\mate#1{#1\sp {\#}}
\catcode`\+=11
\catcode`\@=13

\def\qed{\hspace*{\fill}
\mbox{\hphantom{mm}\rule{0.25cm}{0.25cm}}\\}

\def\bk{{\bf k}}
\def\o{\otimes}
\def\ot{\otimes}
\def\ored{\!\otimes\!}
\def\op{\oplus}
\def\opq{\oplus_q}
\def\oplusq{\oplus_q}
\def\aaa{{(q-q\sp {-1})}}
\def\id{\hbox{\rm 1\hskip -1.5mm 1}}
\def\YB{Yang-Baxter}
\def\yb{{\cal YB}}
\def\hq{{\cal H}_q}
\def\qh{{$q$-Hecke}}
\def\End#1{{\mbox{\rm End}(#1)}}
\def\rsun#1{{\calr_{su(#1)}}}
\def\lll#1{(#1\!\otimes\! \id)}
\def\rrr#1{(\id\!\otimes\! #1)}
\def\cald{{\cal D}}
\def\calc{{\cal C}}
\def\calr{{\cal R}}
\def\cals{{\cal S}}
\def\calu{{\cal U}}
\def\calt{{\cal T}}

\def\calq{{\cal Q}}
\def\cale{{\cal E}}

\def\ox{\overline x}
\def\oy{\overline y}
\def\adjoint#1#2#3#4{#1#2#3\dashv #4}
\def\dual#1#2{{#1}\sp {[#2]}}
\def\trace#1{\mbox{\rm Trace}(#1)}

\long\def\comment#1\endcomment{}
\long\def\cskip#1\endcskip{\hfill\break\par #1\break\par\vskip3.5mm}
\def\lmt{\longmapsto}
\def\glueing{\calr\oplusq\calr'}
\def\Vect{{\mbox{$\bk$-\rm Vect}}}

\def\tp#1#2{{#1}\sp {\otimes#2}}
\def\leftQYBE#1{(#1\o\id)(\id\o#1)(#1\o\id)}
\def\rightQYBE#1{(\id\o#1)(#1\o\id)(\id\o#1)}
\def\leftqybe#1#2#3{(#1\o\id)(\id\o#2)(#3\o\id)}
\def\rightqybe#1#2#3{(\id\o#1)(#2\o\id)(\id\o#3)}
\def\Hom#1#2#3{{\mbox{\rm Hom}_{#1}(#2,#3)}}

\comment Shahn's Personal Macros \endcomment

\def\Bbb#1{{\bbb #1}}
\def\haj#1{{\mathaccent20 {#1}}}
\def\Vhaj{{V\haj{\ }}}

\def\und#1{{\underline {#1}}}

\def\eqn#1#2{\begin{equation}#2\label{#1}\end{equation}}
\def\tens{\mathop{\otimes}}
\def\vect{{\bf t}}\def\vecv{{\bf v}}
\def\veca{{\bf a}}\def\vecx{{\bf x}}\def\vecc{{\bf c}}
\def\vecb{{\bf b}}\def\vecy{{\bf y}}\def\vecd{{\bf d}}
\def\eps{{\epsilon}}
\def\ev{{\rm ev}}
\def\coev{{\rm coev}}
\font\tenbbb=msym10
\font\sevenbbb=msym7
\font\fivebbb=msym5
\newfam\bbbfam \def\bbb{\fam\bbbfam\tenbbb}
\textfont\bbbfam=\tenbbb
\scriptfont\bbbfam=\sevenbbb
\scriptscriptfont\bbbfam=\fivebbb

\begin{document}
\baselineskip 20pt

\begin{center}
\hfill DAMTP/93-20
\end{center}
\vspace{.5in}

\begin{center} {\Large GLUEING OPERATION FOR R-MATRICES, QUANTUM
GROUPS\\ AND
LINK-INVARIANTS OF HECKE TYPE}
\\ \baselineskip 13pt{\ }
\catcode`\@=11
{\ }\\ Shahn MAJID\footnote{SERC Fellow and Fellow of Pembroke
College,
Cambridge}\\ {\ }\\
Department of Applied Mathematics \& Theoretical Physics\\ University
of
Cambridge, CB3 9EW, U.K.\\
e-mail: majid{@}damtp.cambridge.ac.uk\\
{\ }\\
Martin MARKL\\
{\ }\\
Mathematical Institute of the Academy
\\ \v Zitn\'a~25, 115~67 Prague, Czech Republic\\
e-mail: markl{@}csearn.bitnet
\catcode`\@=13
\end{center}

\begin{center}
July 1993\end{center}
\vspace{10pt}
\begin{quote}\baselineskip 13pt
\noindent{\bf Abstract.} We introduce an associative glueing operation
$\oplus_q$ on the space of solutions of the Quantum Yang-Baxter
Equations of
Hecke type. The corresponding glueing operations for the associated
quantum
groups and quantum vector spaces are also found. The former involves
$2\times
2$ quantum matrices whose entries are themselves square or rectangular
quantum
matrices. The corresponding glueing operation for link-invariants is
introduced
and involves a state-sum model with Boltzmann weights determined by the
link
invariants to be glued. The standard $su(n)$ solution, its associated
quantum
matrix group, quantum space and link-invariant arise at once by
repeated
glueing of the one-dimensional case.
\end{quote}
\baselineskip 20pt
\parskip3pt

\section{Introduction}

Matrix representations of the Artin braid group, also called
Yang-Baxter
operators or R-matrices, have been extensively studied in the last
several
years. Each solution leads to interesting algebraic structures such as
quantum
groups~\cite{FRT:lie} and braided
groups~\cite{Ma:exa}\cite{Ma:introm} as well as
(in the nice
cases) to link-invariants. The celebrated Jones polynomial for knots
and links
is of this type and corresponds to the R-matrix for the quantum group
$GL_q(2)$.

In spite of this great activity, the problem still remains how to
obtain such
Yang-Baxter operators. In concrete terms they are matrix solutions
$R\in
M_n\tens M_n$ for some $n$ of the cubic equations
\eqn{QYBE-mat}{ R_{12}R_{13}R_{23}=R_{23}R_{13}R_{12}.}
Here $M_n$ denotes $n\times n$ matrices and $R_{12}=R\tens 1$ in
$M_n\sp {\tens
3}$ etc. These cubic equations are highly-overdetermined and yet have
a rich
variety if solutions. In general, little is known about the moduli
space of
solutions for $n>2$.

We study in this paper part of this moduli space, namely the solutions
of
(\ref{QYBE-mat}) of Hecke type. These are solutions with quadratic
minimal
polynomials for the associated braid representation $\calr$. Without
loss of
generality we can suppose that the minimal polynomial takes the form
\eqn{Hecke-eqn}{(\calr-q)(\calr + q\sp {-1})=0}
for some $q$ which we fix. On the other hand we do not fix any size
$n$, so we
consider all Hecke solutions together of any dimension or infinite
dimensional.
Note that Hecke solutions have been studied in \cite{Gur:alg} but with
other
results than those
here. The novel result in this paper is an associative glueing
operation
$\oplus_q$
for such solutions. This construction has a great many consequences.
First,
the standard solutions associated to the Lie algebra $su(n)$ (which
are known
to be of Hecke type) are understood now as the n-fold $\oplus_q$ of
the trivial
1-dimensional solution $R=(q)$. Thus we answer the question how to
obtain new
R-matrices from old ones in a way that gives the standard non-trivial
R-matrices from something trivial. This is the main result of
Section~2, where
formal definitions (in an operator notation) can be found. We also
express our
$\oplus_q$ operation in categorical terms in this section as some kind
of
direct sum.

Many of the constructions that are usually made with $R$-matrices or
Yang-Baxter operators behave well under our glueing operation. Thus
the quantum
matrix bialgebra $M_q(n)$ (whose localization is the quantum group
$GL_q(n)$)
is built up in a natural way from an iterated glueing operation
$*{}_q$.
Since the construction is associative one has just as well
\[ M_q(m+n)=M_q(m)*{}_q M_q(n),\qquad\forall m,n\ge 1.\]
Equally well we can glue the quantum matrices $A(R)$ coming from any
R-matrices
of Hecke type to obtain new ones. This is the main result of
Section~3. The
general structure of the glueing $A(R)*{}_q A(R')$ here involves the
novel idea
of a $2\times 2$ quantum block matrix whose entries are themselves
quantum
matrices. The off-diagonal blocks here involve the notion of an
$m\times n$
quantum matrix algebra $A(R:R')$, which we also introduce.

In Section~4 we give the same construction of the associated quantum
spaces or
exchange algebras. This setting is the original motivation behind our
construction. Namely, the natural structure behind these quantum
spaces (such
as the celebrated quantum plane $yx=qxy$) is that of a
braided-vector-space.
I.e., they form Hopf algebras in braided categories~\cite{Ma:introm}
with the
braided-coproduct expressing linear addition of vectors. The quantum
plane
lives in the category associated to the $GL_q(2)$ R-matrix. On the
other hand,
there is no canonical tensor product for Hopf algebras in strictly
braided
categories because the usual tensor product construction becomes
tangled up in
the braided case. Our result provides a partial solution to this
problem. At
least in the Hecke case the braided vector space algebras can be
tensor
produced in some way provided their associated braidings are also
glued
according to our $\oplus_q$ construction. Thus
\[\Bbb C_q\sp {m+n}=\Bbb C_q\sp m\otimes_q \Bbb C_q\sp n,\qquad
\forall m,n\ge
1\]
as Hopf algebras in three different braided categories. Equally well
we can
glue quantum spaces associated to any two Hecke R-matrices to obtain a
third
one. The tensor product ${\tens}_q$ needed here is the braided tensor
product
of
$\Bbb Z$-graded algebras along the lines introduced in \cite{Ma:any}.

We turn in Section~5 to the behaviour of quantum traces and ribbon
elements
under our $\oplus_q$ construction. These are needed for the
construction of
link-invariants as well as (much the same thing) to the construction
of braided
categories with duals associated to R-matrices. Some basic facts here
about
such dualisable R-matrices are developed in this section also.

Finally in Section~6 we conclude with a simple state-sum model
description of
the glueing of link-invariants of Hecke type to obtain new ones. The
construction generalizes Kauffman's state model for the Jones
polynomial.
We do not expect to obtain genuinely new invariants (they should all
be
restrictions of the HOMFLY polynomial) but we do obtain a new
associative
operation among them.

\section{Glueing of Yang-Baxter Operators}

All objects in this paper are assumed to be defined over a
fixed field $\bk$ of characteristic zero. Typically $\bk =
\Bbb C$, the field of complex numbers. As a suitable
reference for the conventions we
recommend~\cite{Ma:qua}.

\begin{definition}
\label{YB}
Let $V$ be a vector space. By a \YB\ operator on $V$
we mean an invertible operator $\calr \in \End{V\otimes
V}$ satisfying the following Quantum \YB\ equation (QYBE):
\[
(\calr\o\id)(\id\o\calr)(\calr\o\id)=
(\id\o\calr)(\calr\o\id)(\id\o\calr).
\]
\end{definition}

If $V$ is finite dimensional with a basis $(x_a)$, we may
introduce the matrix $R = (R\sp {a}{}_i{}\sp {b}{}_{j})$ by
\[
\calr(x_i\o x_j) = x_b \o x_a R\sp a{}_i{}\sp b{}_j.
\]
The QYBE written in terms of $R$ then reads as in (\ref{QYBE-mat}).
This is the form familiar in physics so we shall state our results in
this form
as well as in the operator form with which we work.

\begin{definition}
\label{Hecke}
Let $q \in \bk$ be an invertible element. By a $q$-Hecke
operator we mean an element $\calr \in \End{V\o V}$
satisfying
\[
(\calr - q)(\calr + q\sp {-1}) = 0,
\ \mbox{or, equivalently,}
\ \calr\sp 2 -\aaa\calr-\id=0.
\]
\end{definition}

\begin{definition}
1. Denote by $\yb$ the category whose objects are couples $(V,\calr)$,
where
$V$ is a vector space and $\calr$ is a
\YB\ operator on $V$. Morphisms from $(V,\calr)$ to
$(W,\cals)$ in $\yb$ are linear maps $f:V\to W$ such that
$\cals \circ(f\ot f)=(f\ot f)\circ\calr$.

2. Denote by
$\hq$ the category whose objects are the couples
$(V,\calr)$, where $\calr$ is an Yang-Baxter
q-Hecke operator on $V$ and
whose morphisms from $(V,\calr)$ to $(W,\cals)$ are linear
maps $h:V\ot V \to W\ot W$ such that $\cals \circ h =
h\circ \calr$.
\end{definition}

Notice the substantial difference of the
nature of morphisms in $\yb$ and $\hq$. We will see the need for this
later
in the section.

One of the basic constructions of this paper is described in the
following theorem.

\begin{theorem}
\label{glueing}
Let $(X,\calr)$ and $(Y,\calr')$ be elements of $\hq$.
Define their {\em glueing} $\calr\oplus_q\calr' \in \End{(X\oplus Y)\o
(X\oplus
Y)}$ by
\begin{eqnarray*}
&(\calr\oplus_q\calr')|_{X\o X} = \calr,\quad
(\calr\oplus_q\calr')|_{Y \o Y} =
\calr',&
\\
&(\calr\oplus_q\calr')(x\o y) = y\o x,\quad (\calr\oplus_q\calr')(y\o
x)= x\o
y + \aaa
\cdot y\o x&
\end{eqnarray*}
for $x\in X$ and $y\in Y$.
Then $(X\oplus Y,\calr\oplus_q\calr') \in \hq$.
\end{theorem}

\noindent
{\bf Proof.}
We postpone the proof to \ref{glue} where we show that
$\calr\oplusq\calr'$ is a special case of a
much more general construction described in Theorem~\ref{fusion}.
\qed

Under a suitable choice of
conventions, the corresponding matrix $R\oplus_q R' =
((R\oplus_qR')\sp {ab}_{ij})$ can be written in a nice
block form. To be more precise, let $X=\mbox{Span}(x_k;\
1\leq k\leq m)$ and $Y=\mbox{Span}(y_l;\
m+1\leq k\leq m+n)$. If $I=\{kl; 1\leq k,l\leq m+n\}$, then
we can write $I = I_X \cup I_<\cup I_>\cup I_Y$
(the disjoint union), where
\begin{eqnarray*}
&I_X = \{kl \in I;\ k,l\leq m\},\ I_Y = \{kl \in I;\
k,l\geq m+1\},&
\\
&I_< = \{kl \in I;\ k\leq m\ \&\ l\geq m+1\}\ ,\quad
I_>=\{kl \in I;\ k\geq m+1\ \&\ l\leq m\}.&
\end{eqnarray*}
We can order $I$ by requiring that $I_X < I_< < I_> < I_Y$
and that each of the four pieces $I_X$, $I_Y$, $I_<$ and $I_>$
has the standard lexicographic order. Then
\eqn{glue-mat}{
R\oplus_q R'=\left(
\begin{array}{cccc}
R&0&0&0\\\
0&\id&\aaa P&0\\
0&0&\id&0\\
0&0&0&R'
\end{array}
\right).}
Here $(R\oplus_q R')\sp {ab}_{ij}$ is given as the above
matrix entry at row $ab$ and column $ij$ in the order
introduced above, where $R$ (resp.~$R'$) is the matrix
representing $\calr$
(resp.~$\calr'$), $\id$ is the identity matrix and $P$ is the
permutation matrix, $P\sp {ab}_{ij} = \delta\sp a_j\delta\sp b_i$
(Kronecker deltas).

\begin{example}
{\rm
\label{su(n)}
Let $n \geq 1$ and let $E_{ij}$ be, for $1\leq i,j\leq n$,
the $(n\times n)$-matrix given by $(E_{ij})\sp k_l =
\delta\sp k_i\delta\sp j_l$. Then the well-known \YB\ operator $\rsun
n$
related with the semisimple Lie algebra $su(n)$ and the matrix quantum
groups
$GL_q(n)$ and $SL_q(n)$ is given by
the matrix
\[
R_{su(n)}:=
(q\sum_i E_{ii}\o E_{ii}\oplus \sum_{i\not= j}E_{ii}\otimes
E_{jj} +\aaa\sum_{j>i} E_{ij}\o E_{ji})
\]
It can easily be seen that the corresponding operator is a glueing
\[
\rsun{m+n}=\rsun m \oplusq \rsun n,\
\mbox{for $m,n \geq 1$}.
\]
Thus the glueing enables us to construct $\rsun m$
inductively by
\[
\rsun{m+1} = \rsun m \oplusq \rsun 1
\]
from the one-dimensional solution $\rsun 1(1\tens 1) = q 1\tens 1$.
This
description of $\rsun n$ ensures that it possesses the $q$-Hecke
property, as
well as any other properties which hold for $\rsun 1$ and which are
preserved
under $\oplus_q$.
}\end{example}

\begin{example}
{\rm\label{AlexConway}
Let $\calr,\calr' \in \End{\bk\o\bk}$ be the \YB\ $q$-Hecke operators
given by
$\calr(1\o1)=q\cdot(1\o1)$
and ~$\calr'(1\o1)=-q\sp {-1}\cdot(1\o1)$ respectively. Then
$\calr\oplusq\calr'$
is the non-standard \YB\ $q$-Hecke operator
$\calr_{{gl}(1|1)}$ given by the matrix
\[
\left(
\begin{array}{cccc}
q&0&0&0\\\
0&1&q-q\sp {-1} &0\\
0&0&1&0\\
0&0&0&-q\sp {-1}
\end{array}
\right)
\]
}\end{example}

The glueing is a special case of a more general
construction described in the following theorem.

\begin{theorem}
\label{fusion}
Let $X$ and $Y$ be vector spaces and let $\calr :X\o X\to
X\o X$, $\calr' : Y\o Y \to Y\o Y$, $\calu :X\o Y \to Y\o
X$, $\cals :Y\o X\to X\o Y$ and $\calt : Y\o X \to Y \ot X$
be linear maps. Define $\calr\oplus_{\calu,\cals,\calt}\calr' \in
\End{(X\oplus
Y)\o
(X\oplus Y)}$ by
\[
(\calr\oplus_{\calu,\cals,\calt}\calr')|_{X\o X} = \calr,\quad
(\calr\oplus_{\calu,\cals,\calt}\calr')|_{Y\o Y} = \calr',\]
\[(\calr\oplus_{\calu,\cals,\calt}\calr')(x\o y)= \calu(x\o y),\quad
(\calr\oplus_{\calu,\cals,\calt}\calr')(y\o x) = \cals(y\o x)+
\calt(y\o x),
\]
for $x\in X$ and $y\in Y$. This defines a \YB\ operator
on $X\oplus Y$ if and only if both $\calr$ and $\calr'$ are
\YB\ operators, $\calu$ and $\cals$ are invertible, and the
following conditions are satisfied

\vskip2mm
\def\monsterleft#1#2#3#4#5#6#7#8{
\noindent
$(\mbox{\rm #1})$\hfil
$(\id\o#2)(#3\o\id)(\id\o#4)=(#5\o\id)(\id\o#6)(#7\o\id)$\
\mbox{on}\ {$#8$},
\hfil\break\par
\vskip-8pt
}

\def\monsterright#1#2#3#4#5#6#7#8{
\noindent
$(\mbox{\rm #1})'$\hfil
$(#2\o\id)(\id\o#3)(#4\o\id)=(\id\o#5)(#6\o\id)(\id\o #7)$\
\mbox{on}\ {$#8$},
\hfil\break\par
\vskip-8pt
}

\monsterleft{I}{\calr}{\calu}{\calu}{\calu}{\calu}{\calr}
 {\tp X2\ored Y}
\monsterright{I}{\calr'}{\calu}{\calu}{\calu}{\calu}{\calr'}
 {X\ored \tp Y2}
\monsterleft{II}{\cals}{\cals}{\calr}{\calr}{\cals}{\cals}
 {Y\ored \tp X2}
\monsterright{II}{\cals}{\cals}{\calr'}{\calr'}{\cals}{\cals}
 {\tp Y2 \ored X}
\monsterleft{III}{\calu}{\calr}{\cals}{\cals}{\calr}{\calu}
 {X\ored Y\ored X}
\monsterright{III}{\calu}{\calr'}{\cals}{\cals}{\calr'}{\calu}
 {Y\ored X\ored Y}
\monsterleft{IV}{\calr}{\calu}{\calt}{\calt}{\calr}{\calu}
 {X\ored Y\ored X}
\monsterright{IV}{\calr'}{\calu}{\calt}{\calt}{\calr'}{\calu}
 {Y\ored X\ored Y}
\monsterleft{V}{\calt}{\cals}{\calr}{\cals}{\calr}{\calt}
 {Y\ored \tp X2}
\monsterright{V}{\calt}{\cals}{\calr'}{\cals}{\calr'}{\calt}
 {X\ored \tp Y2}
\noindent
$(\mbox{\rm VI})$\hfil
$\rrr{\calr}\lll{\calt}\rrr{\calr}\!=\!
\lll{\calt}\rrr{\calr}\lll{\calt}\!+\!
\lll{\calu}\rrr{\calt}\lll{\cals}\ \mbox{on}\ Y\ored \tp X2$,\hfil
\break
\par
\vskip-8pt
\noindent
$(\mbox{\rm VI})'$\hfil
$\lll{\calr'}\rrr{\calt}\lll{\calr'}\!=\!
\rrr{\calt}\lll{\calr'}\rrr{\calt}\!+\!
\rrr{\calu}\lll{\calt}\rrr{\cals}\ \mbox{on}\ X\ored \tp Y2$.\hfil
\break
\end{theorem}

\noindent
{\bf Proof:}
First, it is immediate to see that
$\calr\oplus_{\calu,\cals,\calt}\calr'$ is
invertible if
and only if both $\calr$, $\calr'$, $\cals$ and $\calu$ are
invertible operators.

We shall verify that $\calq=\calr\oplus_{\calu,\cals,\calt}\calr'$
satisfies
the QYBE on
$\tp{(X\!\op\! Y)}3$, which can be decomposed as
\begin{equation}
\label{decomposition}
\tp X3 \!\op\! (Y\!\o\!\tp X2) \!\op\!
(X\!\o\! Y\!\o\! X) \!\op\! (\tp X2\!\o\! Y) \!\op\!
(X\o\tp Y2) \!\op\! (Y\!\o\! X\!\o\! Y) \!\op\!
(\tp Y2\!\o\! X) \!\op\! \tp Y3.
\end{equation}

On the first component of the previous decomposition we clearly
have
\[
\leftQYBE{\calq}|_{\tp X3}=\leftQYBE{\calr}
\]
and, similarly,
\[
\rightQYBE{\calq}|_{\tp X3}=\rightQYBE{\calr},
\]
therefore QYBE for $\calq$ on the first component
of~(\ref{decomposition}) is equivalent with QYBE for $\calr$.
On the next component of~(\ref{decomposition}) we have
\begin{eqnarray*}
\lefteqn{
\leftQYBE{\calq}|_{(Y\o\tp X2)}=
\leftqybe{\calr}{\cals}{\cals}+}
\\&&
+\leftqybe{\calu}{\calt}{\cals}+\leftqybe{\cals}{\calr}{\calt}
+\leftqybe{\calt}{\calr}{\calt},
\end{eqnarray*}
while
\begin{eqnarray*}
\lefteqn{
\rightQYBE{\calq}|_{(Y\o\tp X2)}=}
\\&&
=\rightqybe{\cals}{\cals}{\calr}+\rightqybe{\calt}{\cals}{\calr}
+\rightqybe{\calr}{\calt}{\calr}.
\end{eqnarray*}
The terms of the above two equations are maps from $(Y\ored \tp X2)$
to
$(Y\ored\tp X2) \op (X\ored Y\ored X) \oplus (\tp X2 \ored Y)$.
Looking at
the component in $(Y\ored\tp X2)$ we get that the QYBE for
$\calq$ imply
\[
\leftqybe{\calu}{\calt}{\cals}+
\leftqybe{\calt}{\calr}{\calt}=
\rightqybe{\calr}{\calt}{\calr},
\]
which is ~(VI). Similarly, the component in $(X\ored Y\ored X)$
gives~(V) and the component in $(\tp X2 \ored Y)$ gives~(II).
Repeating the same discussion for the remaining six
components of~(\ref{decomposition}) we get all the
equations (I)--(VI)' of the theorem. On the other hand, it
also follows immediately from this discussion that the
equations (I)--(VI)' imply the QYBE for $\calq$.
\qed

This provides an extension of our glueing construction
to a more general situation (and is not directly connected with the
Hecke
situation). Using the obvious similar conventions as in
(\ref{glue-mat}), we
can represent the resulting \YB\ operator
by the matrix
\eqn{glue-ust-mat}{
R\oplus_{U,S,T}R'=\left(
\begin{array}{cccc}
R&0&0&0\\\
0&U&T&0\\
0&0&S&0\\
0&0&0&R'
\end{array}
\right)}
where the matrices $R,R',U,S,T$ are matrices representing the
operators in the
theorem. In matrix terms the conditions (I)-(VI) in theorem are
\[ U_{12}U_{13}R_{23}=R_{23}U_{13}U_{12},\
R_{12}S_{13}S_{23}=S_{23}S_{13}R_{12},\
S_{12}R_{13}U_{23}=U_{23}R_{13}S_{12}\]
\[ T_{12}U_{13}R_{23}=U_{23}R_{13}T_{12},\
R_{12}S_{13}T_{23}=S_{23}R_{13}T_{12},\
R_{12}T_{13}R_{23}=T_{23}R_{13}T_{12}+S_{23}T_{13}U_{12}\]
and similarly for the primed equations with respect to $R'$. There is
an
evident mirror-like symmetry between the primed and unprimed equations
in the
theorem.

In the following examples we describe some special choices
of operators satisfying the conditions formulated in
Theorem~\ref{fusion}.

\begin{example}{\rm\
\label{direct_sum}
The following choice of endomorphisms satisfies the
assumptions of Theorem~\ref{fusion}~: $\calr$ and $\calr'$
are arbitrary Yang-Baxter operators, $\calu$ and $\cals$
are given by $\calu(x\o y)=y\o x$,
$\cals(y\o x)=x\o y$, $x\in X$, $y\in Y$, and $\calt=0$. We denote in
this case
the
resulting operator by $\calr\oplus\calr'$
and call it the {\em direct sum} of the
Yang-Baxter operators $\calr$ and $\calr'$.

The equations (I)--(III)' can be verified easily, we have,
for example,
\[
\rrr{\calr}\lll{\calu}\rrr{\calu}(x\o x'\o y)= y\o
\calr(x\o x') = \lll{\calu}\rrr{\calu}\lll{\calr}(x\o x'\o y),
\]
which is (I). The remaining equations (IV)--(VI)' are
trivial, since $\calt=0$.
}\end{example}

\begin{example}{\rm\
\label{special_case}
Let $\calr$ and $\calr'$ be Yang-Baxter operators, $a\in
\bk$, and $\calu$, $\cals$ be invertible operators such
that the conditions (I)-(III)' of Theorem~\ref{fusion} are
satisfied and such that, moreover,\hfil\break
\par\noindent
$(\mbox{\rm VII})$\hfil
$\rrr{\calr\sp 2}=
a\cdot\rrr{\calr}+
\lll{\calu}\lll{\cals}\ \mbox{on}\ Y\o \tp X2$,\hfil
\break
\par
\vskip-10pt
\noindent
$(\mbox{\rm VII})'$\hfil
$\lll{\calr'\sp 2}=
a\cdot\lll{\calr'}+
\rrr{\calu}\rrr{\cals}\ \mbox{on}\ X\o \tp Y2$.\hfil
\break\par\noindent
Then the operators $\calr$, $\calr'$, $\calu$, $\cals$ and
$\calt := a\cdot \id$ satisfy all the equations of
Theorem~\ref{fusion}.

This is almost obvious.
Notice that the conditions (IV)-(V)' are
satisfied trivially, since
$\lll{\calt}=\rrr{\calt}=a\cdot\tp {\id}3$. For the
same reason (VII) implies (VI) and (VII)' implies (VI)'.
}\end{example}

\begin{example}{\rm\
Let $\calr$ be the Yang-Baxter operator on $\bk$ given by
$\calr(1\o1)=q$, $\calr'=\calr$, and
$\calu$, $\cals$ and $\calt$ be defined by
$\calu(1\o1)=qp\sp {-1}$, $\cals = \id$ and
$\calt(1\o1)=q-p\sp {-1}$, for some invertible $p,q\in \bk$.
Then it is easy to verify the
assumptions formulated in Example~\ref{special_case} (with
$a=q-p\sp {-1}$) and our construction
gives a two-parameter solution of the QYBE represented by
the matrix
\[
\left(
\begin{array}{cccc}
q&0&0&0\\\
0&qp\sp {-1}&q-p\sp {-1}&0\\
0&0&1&0\\
0&0&0&q
\end{array}
\right)
\]
}\end{example}

\begin{example}{\rm\
\label{special_case_glueing}
Let $\calr$ and $\calr'$ be Yang-Baxter q-Hecke operators,
$a=\aaa$, $\calu$ be an invertible operator satisfying (I) and
(I)', and $\cals =\calu\sp {-1}$.
Then again this choice satisfies the assumptions formulated
in Example~\ref{special_case}. This is obvious:
under the assumption $\cals = \calu\sp {-1}$, (I)
implies (II) \& (III) and (I)' implies (II)' \& (III)', while
(VII) (resp.~(VII')) becomes the Hecke-type equation for
$\calr$ (resp.~$\calr'$).
}\end{example}

\begin{example}{\rm\
\label{glue}
Let $\calr=\calr'$ be defined by $\calr(1\o1)=q$ and let
$\calu$, $\cals$ be given by $\calu(1\o1)=s\cdot(1\o1)$,
$\cals(1\o1)=s\sp {-1}\cdot(1\o1)$, for some invertible
$q,s\in \bk$. These operators satisfy the conditions
of Example~\ref{special_case_glueing} and our
construction gives another two-parameter solution
represented by the matrix
\[
\left(
\begin{array}{cccc}
q&0&0&0\\\
0&s&q-q\sp {-1}&0\\
0&0&s\sp {-1}&0\\
0&0&0&q
\end{array}
\right)
\]

Another choice of the same type is the following. Let
$(X,\calr),(Y,\calr')\in\hq$, let $\calu$ be given by
$\calu(x\o y)=y\o x$, let $\cals = \calu\sp {-1}$ and let
$\calt = \aaa\cdot \id$. This choice gives the glueing
$\calr\oplusq\calr'$ of Theorem~\ref{glueing}.
}\end{example}

We end this section with a categorical characterization of
some constructions above. Discuss first the `direct sum'
$\calr\op\calr'$ of Example~\ref{direct_sum}.
For $(X,\calr), (Y,\calr')\in
\yb$, the natural inclusions $\iota_1 :X\hookrightarrow
X\op Y$ and $\iota_2 : Y\hookrightarrow
X\op Y$ induce morphisms (denoted by the same symbols)
$\iota_1:(X,\calr)\to (X\op Y,\calr\op\calr')$ and
$\iota_2:(Y,\calr')\to (X\op Y,\calr\op\calr')$ in the
category $\yb$.

\begin{proposition}
Let $(X,\calr)$, $(Y,\calr')$ and $(Z,\calq)$ be elements
of $\yb$ and let
$f_1:(X,\calr)\to (Z,\calq)$ and $f_2:(Y,\calr')\to
(Z,\calq)$ be morphisms in $\yb$ such that
\[
\calq(f_1(x)\o f_2(y))=f_2(y)\o f_1(x),\ \mbox{for all $x\in
X$ and $y\in Y$}.
\]
Then there exists exactly one map $f:(X\op
Y,\calr\op\calr')\to(Z,\calq)$ in $\yb$ such that
$f_i=f\circ \iota_i$, $i=1,2$.
\end{proposition}

\noindent
{\bf Proof.} The map $f$ must necessarily
be given by $f(\iota_1(x))=f_1(x)$ and
$f(\iota_2(y))=f_2(y)$, and if we take it as the definition
of $f$, we easily verify that $\cals \circ (f\o f)=(f\o
f)(\calr\op\calr')$. \qed

This proposition provides the following
categorial characterization of the direct sum $\calr\op\calr'$.
Let $\cald$ be the category whose objects are quadruples of
the form $(V,A,B,\calq)$, with $(V,\calq)\in \yb$, $A$ and
$B$ vector subspaces of $V$, $\calq(A\o A)\subset A\o A$,
$\calq(B\o B)\subset B\o B$, $\calq(a\o b)= b\o a$ and
$\calq(b\o a) = a\o b$, for $a\in A$ and $b\in B$. The
morphisms from $(V,A,B,\calq)$ to $(W,A',B',\calq')$ in
$\cald$ are
morphisms $f :(V,\calq)\to (W,\calq')$ in $\yb$ such that
$f(A)\subset A'$ and $f(B)\subset B'$. Define the functor
$\Delta :\cald \to \yb\times\yb$ by
\[
\Delta(V,A,B,\calq)=(A,\calq)\times(B,\calq).
\]
Then we have the following characterization (compare the
characterization of direct sums as it is
given in~\cite{HilSta:cou}).

\begin{proposition}
\label{direct_sum1}
The functor $\op : \yb \times\yb \to \cald$,
$(X,\calr)\times(Y,\calr')\mapsto (X\op Y,X,Y,\calr\op
\calr')$, is a left adjoint to the functor $\Delta:\cald
\to \yb \times\yb$.
\end{proposition}

\noindent
{\bf Proof.}
The proof is fully routine one, and we add it only
because we do not suppose that this kind of arguments is
commonly known among the readers of this paper.

We must show that, for any $\xi=(V,A,B,\calq)\in \cald$ and
$\eta = (X,\calr)\times(Y,\calr')\in \yb\times\yb$, we have a
functorial isomorphism
\[
\Hom{\cald}{\op \eta}{\xi}\cong
\Hom{\yb\times\yb}{\eta}{\Delta \xi}
\]
or, expanding the expressions,
\[
\Hom{\cald}{(X\op Y,X,Y,\calr\op\calr')}{(V,A,B,\calq)}
\cong \Hom{\yb}{(X,\calr)}{(A,\calq)}\times
\Hom{\yb}{(Y,\calr')}{(B,\calq)}.
\]
But this is almost clear. An element from the left-hand
side Hom-set is given by a linear map $f :X\op Y\to V$. The
maps $f_i := f\circ \iota_i$, $i=1,2$, define the morphisms
(denoted by the same symbols)
$f_1:(X,\calr)\to (A,\calq)$ and $f_1:(Y,\calr')\to
(B,\calq)$ from $\yb$, thus $f_1\times f_2$ is an element of
the right-hand side Hom-set.

On the other hand, two such maps $f_1$ and $f_2$ give rise
to the element $f$ from the left-hand side
Hom-set exactly as it was shown in
Proposition~\ref{direct_sum1}.
\qed

Let us now consider the glueing $\calr\opq\calr'$ in the same
categorical way.
We again
have the natural inclusions $j_1:X\to X\op Y$ and $j_2:Y\to
X\op Y$ which induce, by $\iota_1(x\o x'):= j_1(x)\o
j_1(x')$ and $\iota_2(y\o y'):= j_2(y)\o
j_2(y')$ the morphisms (denoted by the same symbols)
$\iota_1:(X,\calr) \to (X\op Y,\calr\opq \calr')$ and
$\iota_2:(Y,\calr') \to (X\op Y,\calr\opq \calr')$ in $\hq$.
We then have the following proposition.

\begin{proposition}
\label{abstract_nonsense}
Let $(X,\calr)$, $(Y,\calr')$ and $(Z,\calq)$ be elements
of $\hq$, let $f_1:(X,\calr)\to (Z,\calq)$ and
$f_2:(Y,\calr')\to (Z,\calq)$ be morphisms in $\hq$, and let
$\phi: X\o Y\to Z$ be a linear map.
Then there exists exactly one morphism $f: (X\op
Y,\calr\oplus_q\calr')\to (Z,\calq)$ in $\hq$ such that
\[
f\circ \iota_1=f_1,\ f\circ \iota_2=f_2\ ,\quad
f(j_1(x)\o j_2(y))=\phi(x\o y),
\]
for any $x\in X$ and $y\in Y$.
\end{proposition}

\noindent
{\bf Proof.}
We identify, for simplicity, the elements of $X$
(resp.~$Y$) with their images under $j_1$ (resp.~$j_2$) in
$X\op Y$. We already know that
\[
f|_{X\o X}= f_1,\ f|_{Y\o Y} = f_2\ ,\quad f|_{X\o Y}=\phi.
\]
Suppose that $f|_{Y\o X} = \psi$ for some linear map $\psi :Y\o X\to
Z$. The basic equation
\[
\calq\circ f = f\circ (\calr\opq\calr')
\]
is evidently satisfied on $X\o X$ and $Y\o Y$. For $x\in X$
and $y\in Y$ it gives
\[
\calq\circ\phi(x,y)=\psi(y,x)\ ,\quad \calq\circ
\psi(y,x)=\phi(x,y)+ \aaa \cdot\psi(y,x).
\]
We can take the first equation above as the definition of
$\psi$, i.e.~put $\psi=\calq \circ \phi$. The second
equation then reads
\[
\calq\sp 2\circ \phi = \phi - \aaa\cdot\calq\circ\phi
\] and it is clearly a consequence of the Hecke equation
for $\calq$. \qed

Again we can convert the previous example into a categorial
characterization of the glueing. Let $\cale$ be the
category whose objects are couples $((X,\calr),(Y,\calr'))$
of elements of $\hq$ and the morphisms between
$((X,\calr),(Y,\calr'))$ and
$((X_1,\calr_1),(Y_1,\calr'_1))$ in $\cale$ are triples $(f,g,\phi)$,
with $f\in \Hom{\hq}{(X,\calr)}{(X_1,\calr_1)}$,
$g\in \Hom{\hq}{(Y,\calr')}{(Y_1,\calr_1')}$, and $\phi:X\o
Y\to \tp {(X_1\op Y_1)}2$
an arbitrary linear map. Define the functor $\Delta
:\hq \to \cale$ by $\Delta((X,\calr))
=((X,\calr),(X,\calr))$. The proof of the following
proposition is
almost identical with the proof of
Proposition~\ref{abstract_nonsense} and we omit it.

\begin{proposition}
The functor $\opq :\cale \to \hq$ sending
$((X,\calr),(Y,\calr'))\to (X\op Y,\calr\opq\calr')$, is a
left adjoint to the functor $\Delta:\hq\to\cale$ defined
above.
\end{proposition}

\section{Glueing of Quantum Matrices}

We recall that a bialgebra over $\bf k$ is $(A,\Delta,\eps)$ where $A$
is a
unital algebra
and $\Delta:A\to A\tens A$ a co-associative algebra homomorphism
forming a
coalgebra with counit $\eps:A\to {\bf k}$.

Associated to every matrix $R\in M_n\tens M_n$ is a bialgebra $A(R)$
defined as
follows. We take a $n\sp 2$ generators $\vect=(t\sp i{}_j)$ regarded
as a
matrix, and
relations~\cite{FRT:lie} cf~\cite{Dri}
\eqn{FRT}{
R(\vect\tens\id)(\id\tens\vect)=(\id\tens\vect)(\vect\tens\id)R,\quad
{\rm
i.e.,}\quad R\vect_1\vect_2=\vect_2\vect_1R}
in a standard compact notation. The coproduct and counit are
\eqn{mat-coprod}{ \Delta t\sp i{}_j=t\sp i{}_a\tens t\sp a{}_j,\quad
\eps
t\sp i{}_j=\delta\sp i{}_j,\quad{\rm i.e.,}\quad
\Delta\vect=\vect\tens\vect,\quad
\eps\vect=\id.}
This structure is well-known and is commonly called a {\em quantum
matrix algebra}. One
assumes that $R$ is the matrix representing an invertible \YB\
operator in
order to be sure that the quantum matrix algebra has canonical matrix
representations and other nice properties. In this case it is
quasi-commutative
in the sense of a dual-quasitriangular structure induced by $R$
cf~\cite{Ma:seq}.

Further quotienting of these $A(R)$ for standard solutions $R$ gives
the
matrix description of the quantum function algebras ${\cal O}_q(G)$
deforming
the ring of representative functions on the standard semisimple Lie
groups~\cite{FRT:lie}. The standard $\rsun n$ solutions give the
quantum
matrices $M_q(n)$ and the quotient of these by a $q$-determinant
relation gives
the matrix quantum group $SL_q(n)$ or (in a suitable context)
$SU_q(n)$. One
can also localize the $q$-determinant and obtain the matrix quantum
group
$GL_q(n)$. In all cases these algebras are like the ring of functions
on some
`quantum group' from the point of view of non-commutative geometry.

On the other hand, the $A(R)$ bialgebra construction is not limited to
these
standard Yang-Baxter operators and in general need not have any such
determinants etc. Thus for the general theory we are limited to the
quantum
matrix setting. We give in this section a description of $A(R\oplus_q
R')$
where $R{\oplus}_qR'$ as from (\ref{glue-mat}). In order to do this we
will be
led
to introduce the notion of a rectangular quantum matrix algebra. These
include
the quantum matrix algebras above as well as the quantum spaces of the
next
section.

\begin{definition} Let $R\in M_m\tens M_m$ and $R'\in M_n\tens M_n$ be
the
matrices. We define the $m\times n$ {\em rectangular quantum matrix
algebra}
$A(R:R')$ to be the algebra generated by $1$ and $\vect=(t\sp
i{}_{j'})$ for
$1\le
i\le m$ and $1\le j'\le n$ with relations
\[ R(\vect\tens\id)(\id\tens\vect)=(\id\tens
\vect)(\vect\tens\id)R',\quad {\rm
i.e.,}\quad R\vect_1\vect_2=\vect_2\vect_1R'\]
\end{definition}
We use here the same compact notation as above for quantum matrices.
If one
wants to see the relations written completely explicitly in terms of
the
individual generators, they are
\eqn{A(R:R')}{ R\sp i{}_{a}{}\sp k{}_b t\sp a{}_{j'} t\sp
b{}_{l'}=t\sp
k{}_{b'}
t\sp i{}_{a'} R'{}\sp {a'}{}_{j'}{}\sp {b'}{}_{l'},\quad\forall\ 1\le
i,k\le
m,\ 1\le
j',l'\le n}
where summation is over $1\le a,b\le m$ and $1\le {a'},{b'}\le n$. We
use
unprimed indices for the range $1,\cdots,m$ and primed indices for the
range
$1,\cdots,n$. An elementary proposition is the following.

\begin{proposition} Let $R\in M_m\tens M_n$ and $R'\in M_n\tens M_n$
be
matrices and let
\[\veca=(a\sp i{}_j\in A(R)),\quad \vecb=(b\sp i{}_{j'}\in
A(R:R')),\quad
\vecc=(c\sp {i'}{}_j\in A(R':R)),\quad \vecd=(d\sp {i'}{}_{j'}\in
A(R'))\]
denote the generators of the corresponding square and rectangular
quantum
matrices as shown. The indices run in the ranges described in the
conventions
above. Then there are algebra homomorphisms

1. $A(R)\to A(R:R')\tens A(R':R),\qquad a\sp i{}_j\mapsto b\sp
i{}_{a'}\tens
c\sp {a'}{}_j$

2. $A(R')\to A(R':R)\tens A(R:R'),\qquad d\sp {i'}{}_{j'}\mapsto c\sp
{i'}{}_a\tens
b\sp a{}_{j'}$

3. $A(R:R')\to A(R)\tens A(R:R'),\qquad b\sp i{}_{j'}\mapsto a\sp
i{}_a\tens
b\sp a{}_{j'}$

4. $A(R:R')\to A(R:R')\tens A(R'),\qquad b\sp i{}_{j'}\mapsto b\sp
i{}_{a'}\tens
d\sp {a'}{}_{j'}$.
\end{proposition}
{\bf Proof.} The proof is easy in the compact notation. Thus
$A(R:R')\tens
A(R':R)$ is generated by $1$ and $\vecb,\vecc$ which mutually commute.
We have
to show that $\veca\mapsto \vecb\vecc$ is an algebra homomorphism.
Thus
$R\vecb_1\vecc_1\vecb_2\vecc_2=R\vecb_1\vecb_2\vecc_1\vecc_2=\vecb_2\vecb_1
R'\vecc_1\vecc_2=\vecb_2\vecb_1 \vecc_2\vecc_1R=\vecb_2\vecc_2
\vecb_1\vecc_1R$
as required for the first map. The proofs for the other maps are
similar. \qed

This proposition extends the notion of quantum linear algebra as in
\cite{Ma:lin} to the setting of rectangular matrices. The first map
corresponds
to the ability to multiply an $m\times n$ matrix with an $n\times m$
matrix to
obtain an $m\times m$ matrix. Similarly for the second map to obtain
an
$n\times n$ matrix. The third map corresponds to the product of an
$m\times m$
matrix with an $m\times n$ matrix from the left. Similarly for the
fourth map
from the right by an $n\times n$ matrix. These maps are therefore the
rectangular counterparts of the coproduct $\Delta$ which corresponds
to usual
square matrix multiplication. The most general statement which covers
all these
maps as special cases is the following: if $R,R'$ are as above and
$R''\in
M_p\tens M_p$ then there is an algebra homomorphism
\eqn{rect-coprod}{ \Delta_{R,R',R''}:A(R:R'')\to A(R:R')\tens
A(R':R''),\quad
\vect\mapsto \vecb\tens\vecc}
where $\vect$ is a quantum $m\times p$ matrix, $\vecb$ is $m\times n$
and
$\vecc$ is $n\times p$ and the map corresponds to the matrix
multiplication of
the latter two to obtain the former. Moreover, the family of maps {\em
taken
together} are coassociative in the sense
\eqn{rect-coassoc}{ ({\rm id}\tens
\Delta_{R',R'',R'''})\circ\Delta_{R,R',R'''}=(\Delta_{R,R',R''}\tens{\rm
id})\circ\Delta_{R,R'',R'''}}
as a map $A(R:R''')\to A(R:R')\tens A(R':R'')\tens A(R'':R''')$. This
includes
all possible coassociativity conditions arising from associativity of
rectangular
matrix multiplication. For example, it says that $\Delta$ in
(\ref{mat-coprod}) is coassociative and that $A(R:R')$ is a left-right
$A(R)-A(R')$-bicomodule via the last two maps in the proposition. In
summary,
rectangular quantum matrices are not individually bialgebras but they
all fit
together into a weaker `co-groupoid' structure on the entire family.
We can now
present our main result of this section in terms of such objects.

\begin{theorem}\label{glue-A(R)} Let $R\in M_m\tens M_m$ and $R'\in
M_n\tens
M_n$ be the matrices of finite-dimensional q-Hecke Yang-Baxter
operators. Then
$A(R\oplus_q R')$ is the algebra generated by $A(R),A(R:R'),A(R':R)$
and
$A(R')$
as subalgebras, modulo relations between them. We write the matrix
generator of
$A(R\oplus_qR')$ as
\[ \vect=\pmatrix{\veca&\vecb\cr\vecc&\vecd},\quad{\rm where},\quad
\veca=(a\sp i{}_j),\quad \vecb=(b\sp i{}_{j'}),\quad
\vecc=(c\sp {i'}{}_j),\quad \vecd=(d\sp {i'}{}_{j'})\]
where $i,j=1,\cdots m$ and $i',j'=1\cdots n$ and
$\veca,\vecb,\vecc,\vecd$ are
the respective matrix generators of the subalgebras. These are as in
Proposition~3.2 and have explicitly the relations

\noindent (I)$\qquad R\veca_1\veca_2=\veca_2\veca_1R,\quad
R\vecb_1\vecb_2=\vecb_2\vecb_1R',\quad
R'\vecc_1\vecc_2=\vecc_2\vecc_1R,\quad
R'\vecd_1\vecd_2=\vecd_2\vecd_1R'.$

\noindent The relations between them are

\noindent (II)$\qquad R\veca_1\vecb_2=\vecb_2\veca_1,\quad
\vecc_1\veca_2=\veca_2\vecc_1R,\quad
R'\vecc_1\vecd_2=\vecd_2\vecc_1,\quad
\vecd_1\vecb_2=\vecb_2\vecd_1R'$

\noindent (III) $\qquad \vecb_1\vecc_2=\vecc_2\vecc_1,\quad
\veca_1\vecd_2-\vecd_2\veca_1=(q\sp {-1}-q)P\vecc_1\vecb_2$

\noindent where $P$ is the permutation matrix. Moreover, the matrix
coalgebra
structure in these terms is
\[ \Delta\veca=\veca\tens\veca +\vecb\tens\vecc,\quad
\eps\veca=\id,\quad
\Delta\vecb=\vecb\tens\vecd +\veca\tens\vecc,\quad \eps \vecb=0\]
\[ \Delta\vecc=\vecc\tens\veca +\vecd\tens\vecc,\quad \eps
\vecc=0,\quad
\Delta\vecd=\vecd\tens\vecd +\vecc\tens\vecb,\quad\eps \vecd=\id\]
\end{theorem}
{\bf Proof.} We compute the structure of $A(R\oplus_q R')$ from
(\ref{FRT})
which in explicit component form reads
\[ (R\oplus_q R')\sp I{}_{A}{}\sp K{}_B t\sp A{}_{J} t\sp B{}_{L}=t\sp
K{}_{B}
t\sp I{}_{A} (R\oplus_q R')\sp {A}{}_{J}{}\sp {B}{}_{L},\quad\forall\
1\le
I,J,K,L\le
m+n\]
where the summed indices also run from $1,\cdots,m+n$. As in
(\ref{glue-mat})
we break the
range into $1,\cdots,m$ which we represent with lower case labels and
the rest
which we represent by primed
labels running $1,\cdots,n$. With these conventions the relations
are of 16 types according as the four free labels are each in their
upper or
lower
ranges. In evaluating these we must sum over both types of label so
each
equation has four possible
terms on each side. Note now that the
structure of $R\oplus_q R'$ in (\ref{glue-mat}) is such that there are
only
five types of non-zero
entries,
\[ (R\oplus_q R')\sp i{}_j{}\sp k{}_l=R\sp i{}_j{}\sp k{}_l,\quad
(R\oplus_q
R')\sp {i'}{}_{j'}{}\sp {k'}{}_{l'}=R'{}\sp {i'}{}_{j'}{}\sp
{k'}{}_{l'}\]
\[ (R\oplus_q R')\sp i{}_j{}\sp {k'}{}_{l'}=\delta\sp i{}_j\delta\sp
{k'}{}_{l'},\quad
(R\oplus_q R')\sp {i'}{}_{j'}{}\sp {k}{}_{l}=\delta\sp
{i'}{}_{j'}\delta\sp
{k}{}_{l},\quad
 (R\oplus_q
R')\sp i{}_{j'}{}\sp {k'}{}_{l}=(q-q\sp {-1})\delta\sp i{}_l\delta\sp
{k'}{}_{j'}.\]
We also write the entries of $t\sp I{}_J$ in the block form stated and
write
$Q=R\oplus_q R$ for brevity. Then we evaluate our sixteen relations as

$(ikjl):\quad Q\sp {i}{}_{a}{}\sp {k}{}_{b} t
\sp {a}{}_{j} t\sp {b}{}_{l}=t\sp
{k}{}_{b}
t\sp {i}{}_{a} Q{}\sp {a}{}_{j}{}\sp {b}{}_{l}\quad {\rm i.e.,\ }
R\veca_1\veca_2=\veca_2\veca_1R$

$(ikjl'):\quad Q\sp i{}_{a}{}\sp k{}_b t\sp a{}_{j} t\sp b{}_{l'}
=t\sp k{}_{b'}
t\sp i{}_{a} Q{}\sp {a}{}_{j}{}\sp {b'}{}_{l'}\quad {\rm i.e.,\ }
R\veca_1\vecb_2=\vecb_2\veca_1$

$(ikj'l):\quad Q\sp i{}_{a}{}\sp {k}{}_{b} t\sp a{}_{j'} t\sp
b{}_{l}=t\sp
k{}_{b}
t\sp i{}_{a'} Q{}\sp {a'}{}_{j'}{}\sp {b}{}_{l}+t\sp k{}_{b'}
t\sp {i}{}_{a} Q{}\sp {a}{}_{j'}{}\sp {b'}{}_{l}\quad {\rm i.e.,\ }
R\vecb_1\veca_2=\veca_2\veca_1+\vecb_2\veca_1 (q-q\sp {-1})P$

$(ikj'l'):\quad Q\sp {i}{}_{a}{}\sp {k}{}_{b} t\sp {a}{}_{j'} t\sp
{b}{}_{l'}=t\sp {k}{}_{b'}
t\sp {i}{}_{a'} Q{}\sp {a'}{}_{j'}{}\sp {b'}{}_{l'}\quad {\rm i.e.,\ }
R\vecb_1\vecb_2=\vecb_2\vecb_1R'$

$(ik'jl):\quad Q\sp {i}{}_{a}{}\sp {k'}{}_{b'} t\sp {a}{}_{j}
t\sp {b'}{}_{l}+Q\sp {i}{}_{a'}{}\sp {k'}{}_{b} t\sp {a'}{}_{j} t\sp
{b}{}_{l}=t\sp {k'}{}_{b}
t\sp {i}{}_{a} Q{}\sp {a}{}_{j}{}\sp {b}{}_{l}\quad {\rm i.e.,\ }
\veca_1\vecc_2+(q-q\sp {-1})P\vecc_1\veca_2=\vecc_2\veca_1R$

$(ik'jl'):\quad Q\sp {i}{}_{a}{}\sp {k'}{}_{b'} t\sp {a}{}_{j}
t\sp {b'}{}_{l'}+Q\sp {i}{}_{a'}{}\sp {k'}{}_{b} t\sp {a'}{}_{j} t\sp
{b}{}_{l'}=t\sp {k'}{}_{b'}
t\sp {i}{}_{a} Q{}\sp {a}{}_{j}{}\sp {b'}{}_{l'}\quad {\rm i.e.,\ }
\veca_1\vecd_2+(q-q\sp {-1})P\vecc_1\vecb_2=\vecd_2\veca_1$

$(ik'j'l):\quad Q\sp {i}{}_{a}{}\sp {k'}{}_{b'} t\sp {a}{}_{j'}
t\sp {b'}{}_{l}+Q\sp {i}{}_{a'}{}\sp {k'}{}_{b} t\sp {a'}{}_{j'} t\sp
{b}{}_{l}=t\sp {k'}{}_{b'}
t\sp {i}{}_{a} Q{}\sp {a}{}_{j'}{}\sp {b'}{}_{l}+t\sp {k'}{}_{b}
t\sp {i}{}_{a'} Q{}\sp {a'}{}_{j'}{}\sp {b}{}_{l}$

$\hphantom{(ik'j'l):\quad\ }{\rm i.e.,\ }
\vecb_1\vecc_2+(q-q\sp {-1})P\vecd_1\veca_2=(q-q\sp {-1})
\vecd_2\veca_1P+\vecc_2\vecb_1$

$(ik'j'l'):\quad Q\sp {i}{}_{a}{}\sp {k'}{}_{b'} t\sp {a}{}_{j'}
t\sp {b'}{}_{l'}+Q\sp {i}{}_{a'}{}\sp {k'}{}_{b} t\sp {a'}{}_{j'}
t\sp {b}{}_{l'}=t\sp {k'}{}_{b'}
t\sp {i}{}_{a'} Q{}\sp {a'}{}_{j'}{}\sp {b'}{}_{l'}$

$\hphantom{(ik'j'l'):\quad\ }{\rm i.e.,\ }
\vecb_1\vecd_2+(q-q\sp {-1})P\vecd_1\vecb_2=\vecd_2\vecb_1R'$

and eight similar relations. The $(ikjl),(ikjl'),(ikj'l'),(ik'jl')$
relations
here are as stated while the $(ikj'l)$ relation is equivalent to the
$(ikjl')$
one because $R$ is $q$-Hecke. In matrix terms this means
$R_{21}R=1+(q-q\sp {-1})PR$ where $P$ is the permutation matrix.
Remembering
that
our suffices in the compact notation relate to the position in a
matrix tensor
product, the action by conjugation by $P$ is to permute the relevant
suffices.
Thus $PRP=R_{21}$ etc. as needed in this computation. The action of
$P$ also
means that the $(ik'j'l)$
also simplifies to $\vecb_1\vecc_2=\vecc_2\vecb_1$ while the Hecke
property for
$R$ and $R'$ means in the same way as above that $(ik'jl)$ and
$(ik'j'l')$ are
equivalent to $\vecc_1\veca_2=\veca_2\vecc_1R$ and
$\vecd_1\vecb_2=\vecb_2\vecd_1R'$ respectively. The remaining eight
relations
are reduced in the same way and give the three remaining relations in
(I)-(II).
The form of the coproduct is from $\Delta t\sp I{}_J=t\sp I{}_A\tens
t\sp A{}_J=t\sp I{}_a\tens t\sp a{}_J+t\sp I{}_{a'}\tens t\sp
{a'}{}_J$ giving
four equations
according to the range of $I,J$. The counit is immediate.
\qed

We can think of this theorem as defining a glueing construction
$A(R)*{}_qA(R')$ for quantum matrices consisting of viewing them on
the
diagonal of a $2\times 2$ block and adjoining the relevant
off-diagonal
rectangular matrices. In the present case the $2\times 2$ blocks obey
a blocked
version of the usual $M_q(2)$ algebra. Thus,

\begin{example}{\rm Let $R=(q)=R'$ the one-dimensional q-Hecke
Yang-Baxter
operator. Then $a,b,c,d$ are $1\times 1$ matrices of generators,
relations (I)
are empty and (II)-(III) reduce to the standard commutation relations
\[ qab=ba,\quad ca=acq,\quad qcd=dc,\quad db=dbq,\quad bc=cb,\quad
ad-da=(q\sp {-1}-q)cb\]
for the $2\times 2$ quantum matrices $M_q(2)$. This fits with
Example~\ref{su(n)} in Section~2.}
\end{example}

\begin{example}{\rm Let $R=(q)$ and $R'=(-q\sp {-1})$ as in
Example~\ref{AlexConway} and $q\sp 2\ne -1$. Then $a,b,c,d$ are
$1\times 1$
matrices and obey the
relations
\[ b\sp 2=0,\ c\sp 2=0 \]
\[ qab=ba,\quad ca=acq,\quad cd=-qdc,\quad -qdb=bd,\quad bc=cb,\quad
ad-da=(q\sp {-1}-q)cb\]
for the non-standard $2\times 2$ quantum matrices $M_q(1|1)$ as
computed in \cite{JGW:new}. The first two come from the middle two of
(I) as
$qbb=bb
(-q\sp {-1})$ and $qcc=cc(-q\sp {-1})$ and the remainder from
(II)--(III).}
\end{example}

It is easy to see from the structure in Theorem~\ref{glue-A(R)} that
setting
$\vecb=\vecc=0$ gives a bialgebra surjection
\eqn{diag-surj}{ A(R\oplus_q R')\to A(R)\tens A(R').}
Thus, from (III) we see that $\veca$ and $\vecd$ commute in the
quotient as
they should in $A(R)\tens A(R')$ while the other relations either
become empty
or those of $A(R),A(R')$. This map corresponds to the usual
block-diagonal
embedding of $M_m\times M_n\subset M_{n+m}$ as semigroups. Thus
provided we use
our $\oplus_q$ construction when blocking quantum matrices into bigger
ones, we
have a coherent quantum linear algebra.

For a full understanding of this quantum linear algebra one must
understand not
only multiplication but addition, to which we turn in the next section
at least
for the $1\times n$ or $m\times 1$ cases.

\section{Glueing of Quantum-Braided Vector Spaces}

We have introduced rectangular quantum matrix algebras $A(R:R')$ in
the last
section. When $R'=R$ we have the usual square quantum matrices $A(R)$.
On the
other hand $A(R:\lambda)$ with $\lambda\in {\bf k}$ is the algebra
with
generators and relations
\eqn{vect-alg}{ R\vecv_1\vecv_2=\vecv_2\vecv_1\lambda,\qquad
\vecv=(v\sp i).}
This is the algebra of (left-covariant) quantum vectors and denoted
$V_L(R)$.
Likewise it is clear that $A(\lambda:R)$ is the algebra with
generators and
relations
\eqn{covect-alg}{
\lambda\vecx_1\vecx_2=\vecx_2\vecx_1R,\qquad\vecx=(x_i).}
This is the algebra of (right-covariant) quantum covectors and denoted
$\Vhaj(R)$.

The usual quantum planes $\Bbb C_q\sp n$ are of this type with $R$ the
matrix
representing the operator $\rsun n$ in Example~\ref{su(n)}. The
constructions
work best in the case where $R$ are $q$-Hecke \YB\ operators for in
this case
one has a braided-coproduct, counit and braiding~\cite{Ma:poi}
\eqn{bra-covec}{ \Vhaj(R)=A(R:q),\quad \und\Delta x_i=x_i\tens
1+1\tens
x_i,\quad \und\eps x_i=0,\qquad \Psi(\vecx_1\tens\vecx_2)=q\,
\vecx_2\tens\vecx_1
\, R}
There is also a braided-antipode $\und S x_i=-x_i$. The notion of
braided-Hopf
algebras was introduced by
one of the authors and we refer to the review~\cite{Ma:introm} for
details. The
main difference from the usual axioms is that $\und\Delta: \Vhaj(R)\to
\Vhaj(R)\und\tens\Vhaj(R)$ where $\und\tens$ is not the usual
commuting tensor
product but the braided tensor product algebra
\eqn{btensalg}{ (a\tens b)(c\tens d)=a\Psi(b\tens c)d,\quad \forall
a,b,c,d\in
\Vhaj(R).}
Here $\Psi$ and the other maps extend to products of the generators in
a
natural way. For a braided-categorical
setting (which we defer to the next section) one needs $R$ here to be
invertible so that $\Psi$ is invertible. On the other hand this is not
essential in our direct algebraic application here.

By allowing ourselves these slightly more general braided-coproduct
structures
we see that we can express via $\und\Delta$ the addition of two
covectors to
obtain a new covector, provided we remember the braiding $\Psi$. This
is an
important construction with many corollaries. For example, one obtains
at once
differential calculi on the quantum planes by making infinitesimal
additions~\cite{Ma:fre}.

In this section we compute $\Vhaj(R\oplus_q R)$. There is a strictly
analogous
theory for vectors also. Let us note first that all our algebras have
quadratic
relations and hence are naturally $\Bbb Z$-graded by the degree of the
generators. Now just as there is a standard $\Bbb Z_2$-graded tensor
product of
super-algebras, there is a natural braided tensor product of $\Bbb
Z$-graded
algebras.

\begin{definition}\label{q-tensalg} cf~\cite{Ma:any} Let $A,B$ be
$\Bbb
Z$-graded algebras over
${\bf k}$ and let $q\in {\bf k}$. We define the associative $\Bbb
Z$-graded
braided tensor product algebra $A{\tens}_q B$ as $A\tens B$ with
product
\[ (a\tens b)(c\tens d)=a\Psi_{\Bbb Z}(b\tens c)d,\quad \Psi_{\Bbb
Z}(b\tens c)
=q\sp {|b||c|}c\tens b,\qquad \forall a,c\in A,\ b,d\in B\]
on homogeneous elements of degree $|\ |$.
\end{definition}
The underlying braided tensor product construction here and in
(\ref{btensalg})
has been introduced by one
of the authors as part of the theory of braided groups. See
\cite{Ma:introm}
for details. In the present case it is trivial to verify associativity
directly.

\begin{proposition}\label{glue-V(R)} Let $R\in M_m\tens M_m$ and
$R'\in
M_n\tens M_n$ be the matrices of \qh\ \YB\ operators. Then
$\Vhaj(R\oplus_q R')=\Vhaj(R){\tens}_q\Vhaj(R')$, the $\Bbb Z$-graded
braided
tensor product of $\Vhaj(R)$ and $\Vhaj(R')$.

Explicitly, we write the covector generator of $\Vhaj(R\oplus_q R')$
as
$(\vecx,\vecy)$ where $\vecx=(x_i)$ and $\vecy=(y_{i'})$ are the
respective
covector generators of the subalgebras and have the relations

(I) $\qquad q\vecx_1\vecx_2=\vecx_2\vecx_1R,\quad
q\vecy_1\vecy_2=\vecy_2\vecy_1R'$

\noindent as above. The $\Bbb Z$-graded braided tensor product is
equivalent to
the additional cross relations

(II) $\qquad \vecy_1\vecx_2=q\vecx_2\vecy_1$.

Moreover, the braided-coproduct restricts to those of
$\Vhaj(R),\Vhaj(R')$
which appear
as sub-braided-Hopf algebras.
\end{proposition}
{\bf Proof.} The Definition~4.1 of the braided-tensor product
for $\Bbb Z$-graded algebras immediately gives the
commutation relations (II) between the factors. Conversely, these
commutation
relations can be used to put all the $\vecx$ generators to the left of
the
$\vecy$ generators, from which it follows that this description is
equivalent
to the form of $\tens_q$ in Definition~4.1. To compute
$\Vhaj(R\oplus_q
R')$ directly we use the same conventions for the range of indices as
in the
proof of
Theorem~\ref{glue-A(R)}, so our relations are explicitly
\[ qx_Jx_L=x_Bx_A(R\oplus_qR')\sp A{}_J{}\sp B{}_L,\quad \forall 1\le
J,L\le
m+n.\]
We have four types of equation according to the range of $(JL)$ and in
each one
we have
up to four terms on each side according to the breakdown of the range
of
$(AB)$. Remembering the explicit
form of $Q=R\oplus_qR'$ from Theorem~\ref{glue-A(R)} and denoting the
lower and
upper ranges by $\{x_i\}=\vecx$ and $\{x_{j'}\}=\vecy$, we have

\noindent $(jl):\quad qx_jx_l=x_bx_a Q\sp a{}_j{}\sp b{}_l,\quad {\rm
i.e.,\ }
q\vecx_1\vecx_2=\vecx_2\vecx_1R$

\noindent $(jl'):\quad qx_jx_{l'}=x_{b'}x_a Q\sp a{}_j{}\sp
{b'}{}_{l'},\quad
{\rm
i.e.,\ } q\vecx_1\vecy_2=\vecy_2\vecx_1$

\noindent $(j'l):\quad qx_{j'}x_{l}=x_{b}x_{a'}
Q\sp {a'}{}_{j'}{}\sp {b}{}_{l}+x_{b'}x_a Q\sp a{}_{j'}{}\sp
{b'}{}_{l} ,\quad
{\rm i.e.,\
} q\vecy_1\vecx_2=\vecx_2\vecy_1+\vecy_2\vecx_1(q-q\sp {-1})P$

\noindent $(j'l'):\quad qx_{j'}x_{l'}=x_{b'}x_{a'}
Q{}\sp {a'}{}_{j'}{}\sp {b'}{}_{l'},\quad {\rm i.e.,\ }
q\vecy_1\vecy_2=\vecy_2\vecy_1R'$

as required. The third equation here is redundant.

The braided-coproduct of $\Vhaj(R\oplus_q R')$ is by definition the
linear one
as in (\ref{bra-covec}) on the generators but (as explained) it
extended to
products of the generators as an algebra homomorphism to the braided
tensor
product $\Vhaj(R\oplus_q R')\und\tens \Vhaj(R\oplus_q R')$. The latter
is
defined as in (\ref{btensalg}) with the operator $\Psi$ induced by the
\YB\
operator $\calr\oplus_q\calr'$. But from its definition in
Theorem~\ref{glueing} we know that this operator restricts on the
generators
$x_i$ and $y_{i'}$ to $\calr$ and $\calr'$ respectively. Likewise for
their
extensions to products of the generators. Hence the braided-coproduct
$\und\Delta$ restricted to $\Vhaj(R)$ and $\Vhaj(R')$ coincides with
their
braided-coproducts whose extension to products is defined via
$\calr,\calr'$.
\qed

This proposition was the original motivation for the $\oplus_q$
construction.
Namely, one can ask how to build up higher-dimensional quantum space
algebras
by tensor product of lower-dimensional ones, corresponding from the
point of
view of non-commutative geometry to the usual cartesian product.
Unfortunately,
in any fixed braided category, there is no canonical notion of tensor
product
of braided-Hopf algebras or braided-bialgebras. Our answer is that
each of our
quantum spaces comes with its own braiding $\Psi$ as determined from
its
associated Yang-Baxter matrix $R$. At least in the $q$-Hecke setting
one can
$q$-tensor product them, providing their associated braidings are also
glued by
the $\oplus_q$ construction of Section~2.

\begin{example}{\rm Let $R,R'$ represent the \YB\ operators $\rsun m$
and
$\rsun n$
respectively, as in Example~\ref{su(n)}. The associated
quantum-covector
algebras $\Bbb C_q\sp m$ and $\Bbb C_q\sp n$ are given by
\[ x_jx_i=qx_ix_j,\quad\forall 1\le i<j\le m;\quad y_{j'}y_{i'}=q
y_{i'}
y_{j'},\quad\forall 1\le {i'}<{j'}\le n.\]
Their ${\tens}_q$ is of the same form with covector generator
$(\vecx,\vecy)$
and
similar relations because our ordering of the basis vectors is with
all the
primed generators $y_{{i'}}$ appearing as $x_{m+{i'}}$ in $\Bbb C_q\sp
{m+n}$
and
hence with labels greater than the unprimed ones. Thus
\[ \Bbb C_q\sp {m+n}=\Bbb C_q\sp m{\tens}_q\Bbb C_q\sp n\]
as required from Example~\ref{su(n)} and the above proposition.}
\end{example}

\begin{example}{\rm Let $R=(q)$ and $R'=(-q\sp {-1})$ be the two
1-dimensional
\qh\
\YB\ operators as in Example~\ref{AlexConway}. Then $\Vhaj(R\oplus_q
R')$ is
the algebra
\[ qxx=xxq,\quad qyy=-q\sp {-1}yy,\quad yx=qxy.\]
The first relation is empty and the second is $y\sp 2=0$ if $q\sp 2\ne
-1$. The
remaining relation comes from the braided tensor product ${\tens}_q$.
This is
the
usual non-standard plane associated to the R-matrix in
Example~\ref{AlexConway}.}
\end{example}

We have described here the so-called `bosonic' covectors which appear
in our
general setting as the $1\times n$ matrices $A(q:R)$ and studied how
it behaves
under
$\oplus_q$ in the second input. There are also the so-called
`fermionic'
covectors $A(-q\sp {-1}:R)$ in our
notation. Their behaviour under $\oplus_q$ is the same as in
Proposition~\ref{glue-V(R)} with $q$ replaced everywhere by $-q\sp
{-1}$.
Likewise
there are the analogous vector cases $A(R:q)$ and $A(R:-q\sp {-1})$.
All of
these
important cases can be collected into the following general statement
about the
behaviour of $A(R:R')$ under $\oplus_q$ in each input.

\begin{proposition}\label{glue-A(R:R')} Let $R\in M_m\tens M_m$,
$R'\in
M_n\tens M_n$ and $R''\in M_p\tens M_p$ be the matrices of \qh\ \YB\
operators.
Then

1. $A(R:R'\oplus_q R'')=A(R:R'){}_R\und\tens A(R:R'')$

2. $A(R\oplus_q R':R'')=A(R:R'')\und\tens_{R''}A(R':R'')$

where ${}_R\und\tens$ and $\und\tens_{R''}$ are the braided tensor
product
algebras defined by the cross relations
\[ \vecb\in A(R:R'),\ \vecc\in A(R:R''),\quad
R\vecb_1\vecc_2=\vecc_2\vecb_1\]
\[ \vecb\in A(R:R''),\ \vecc\in A(R':R''),\quad
\vecc_1\vecb_2=\vecb_2\vecc_1
R''\]
respectively.
\end{proposition}
{\bf Proof.} The strategy for $A(R:R'\oplus_q R'')$ is the same as in
Proposition~\ref{glue-V(R)} with the role of $q$ replaced by $R$.
Explicitly
decomposing $R'\oplus_q R''$ along the lines in the proof of
Theorem~3.3 and
putting into the relations
\[ R\sp i{}_a{}\sp k{}_b t\sp a{}_J
t\sp b{}_L=t\sp k{}_Bt\sp i{}_A(R'\oplus_qR'')\sp A{}_J{}\sp
B{}_L,\quad
\forall 1\le i,k\le
m,\quad \forall 1\le J,L\le n+p.\]
for the $m\times (n+p)$ quantum matrices $\vect=(\vecb,\vecc)$, we
obtain the
four equations
\[ R\vecb_1\vecb_2=\vecb_2\vecb_1R',\qquad
R\vecb_1\vecc_2=\vecc_2\vecb_1\]
\[ R\vecc_1\vecb_2=\vecb_2\vecc_1+\vecc_2\vecb_1(q-q\sp {-1})P,\quad
R\vecc_1\vecc_2=\vecc_2\vecc_1 R''.\]
The third relation here reduces to the second just because $R$ is \qh\
so that
$R_{21}R=1+(q-q\sp {-1})PR$ and since
$P\vecc_2\vecb_1=\vecc_1\vecb_2P$. Recall that $P$ always denotes the
relevant
permutation operator in a matrix tensor product. The second relation
is
expresses the non-commutativity in the braided tensor product algebra
${}_R\und\tens$ defined as in (\ref{btensalg}) and
Definition~\ref{q-tensalg}
but with braiding
\[ \Psi(\vecc_1\tens\vecb_2)=R_{21}\vecb_2\tens\vecc_1.\]
The proof for $A(R\oplus_qR':R'')$ is entirely similar with
decomposition
$\vect=\pmatrix{\vecb\cr\vecc}$ of an $(m+n)\times p$ quantum matrix.
\qed

One also has a braided addition law for all quantum matrices $A(R:R')$
when
$R,R'$ are the matrices of \qh\ \YB\ operators,
\[ A(R:R'):\qquad
\und\Delta_{R,R'}\vect=\vect\tens1+1\tens\vect,\quad
\und\eps{\vect}=0,\quad\Psi(\vect_1\tens\vect_2)=R_{21}\vect_2
\tens\vect_1R'\]
forming a braided-Hopf algebra. This precisely generalizes the
addition law
(\ref{bra-covec}) from $1\times n$ to a general $m\times n$ quantum
matrix. For
example it means that the usual square quantum matrix algebra $A(R)$
has this
linear coproduct or `coaddition' $\und\Delta$ provided one uses the
notion of a
braided-Hopf algebra. The coaddition is codistributive with respect to
the
product of quantum matrices in the sense
\[ ({\rm
id}\tens\cdot)\circ\tau_{23}\circ(\Delta_{R,R',R''}\tens
\Delta_{R,R',R''})\circ\und\Delta_{R,R''}=(\und\Delta_{R,R'}\tens{\rm
id})\circ\Delta_{R,R',R''}\]
\[ (\cdot\tens {\rm
id})\circ\tau_{23}\circ(\Delta_{R,R',R''}\tens
\Delta_{R,R',R''})\circ\und\Delta_{R,R''}=({\rm id}\tens
\und\Delta_{R,R'})\circ\Delta_{R,R',R''}\]
where $\tau_{23}$ denotes usual transposition in the second and third
positions. For example, $A(R)$ becomes a `quantum ring' with both its
usual
comultiplication and the above coaddition
in a distributive way. It is easy to see that for trivial $R$ these
notions are
the usual notions for the ring of matrices, expressed in terms of the
algebra
of functions on them.

These remarks about addition of quantum matrices are not directly
related to
our glueing construction and will be developed elsewhere. The
coaddition is
however, compatible with Proposition~\ref{glue-A(R:R')} in the sense
that the
factors appear as sub-braided-Hopf algebras. This provides one more
reason to
be interested in the quantum matrices $A(R:R')$ and their glueing.
Clearly, we
obtain Proposition~4.2 as a corollary from this point of view. On the
other
hand the quantum-braided planes are the most well-known of these
algebras and
of independent interest, and for this reason we emphasized them in the
above.

\section{Some other properties of the glueing}

We show in this section that our $\oplusq$ operation is
well-behaved with respect to a number of constructions
related to duals and link invariants. If $\rsun 1$ has
these properties, it follows for example that $\rsun n$
given by the iterated glueing of $\rsun 1$ (see
Example~\ref{su(n)}), has
automatically these properties, too. Many other properties
of $\rsun n$ can similarly be understood in this way as
arising from the naturality of $\oplusq$.

Let us introduce first some standard notation and
terminology. These notions make sense in any monoidal category, though
we
shall be interested here only in full subcategories of the category
${\bf
k}$-Vect of
finite dimensional vector spaces with its usual tensor product and
associativity. We recall
that a general monoidal category consists of a category $\calc$ and a
functor
$\tens:\calc\times\calc\to\calc$ with properties analogous to those
for ${\bf
k}$-Vect. There is a unit object $\und 1$ for the tensor product, and
related
unity and associativity morphisms which we suppress. For ${\bf
k}$-Vect we have
$\und 1=\bk$.

In a monoidal category, there is a notion of the {\em left dual} of an
object
$V$ (see for example \cite{Del:cat}). This is an object $V\sp *$ and
morphisms
$\coev_V :\und 1 \to V\ot V\sp *$ (the coevaluation) and
$\ev_V:
V\sp *\ot V\to \und 1$ (the evaluation) satisfying
\[
(\ev_V\ot{\rm id})({\rm id}\ot\coev_V)= {\rm id}\ ,\quad
({\rm id}\ot\ev_V)(\coev_V\ot{\rm id})={\rm id}.
\]
We will use the shorthand
$\adjoint{}{}{V\sp *}V$ to indicate this situation.
Sometimes we say also that the pair $(\ev,\coev)$ is an
{\em adjunction} between $V\sp *$ and $V$. A monoidal category for
which every object has a left dual is
then called {\em rigid\/}.

A typical example is the category ${\bf k}$-Vect with $V\sp
*=\mbox{Hom}(V,\bk)$ and usual
evaluation. Notice also that for a basis $(x_1,\ldots,x_n)$ of $V$,
there
always exist the {\em dual basis\/} $(x\sp *_1,\ldots,x\sp *_n)$
characterized by the property that $\coev(1)=\sum x_i\otimes
x_i\sp *$ and $\ev(x\sp *_i\otimes x_j)= \delta_{ij}$
(the Kronecker delta).

Suppose we have a morphism $f:V\to W$ and
two adjunctions $\adjoint{}{}{V\sp *}{V}$ and
$\adjoint{}{}{W\sp *}{W}$. Then there is an adjoint or
{\em
dual} morphism $f\sp *:W\sp *\to V\sp *$ defined by
\[ f\sp *=(\ev_W\tens{\rm id})){\rm id}\tens f\tens{\rm id})({\rm
id}\tens\coev_V).\]
This dualization via $\ev,\coev$ can be usefully generalized to the
case of a
morphism $f:X\ot V\to W\ot Y$ for any objects $X,Y$. In this case we
follow
\cite{Lyu:the} and denote the generalized adjoint morphism by $\mate
f:W\sp
*\ot X\to Y\ot V\sp *$. It is defined in the analogous way as
\[
\mate f = (\ev_W\ot {\rm id}\sp 2)({\rm id}\ot
f\ot{\rm id})({\rm id}\sp 2\ot\coev_V).
\]

Finally, we will need the notion of {\em braided monoidal category}.
This
notion was formally introduced in category
theory in \cite{JoyStr:bra} and also
arises in the representation theory of quantum groups. A braided
monoidal
category is by definition a monoidal category $\calc$ equipped with a
natural
isomorphism $\Psi:-\tens-\to-\tens\sp {\rm op}-$, where
the latter is the opposite
tensor product. This $\Psi$ is called the quasisymmetry or {\em
braiding} and
is required to obey some properties analogous to those for the usual
transposition of vector spaces, namely
\[
\Psi_{V,W\tens Z}=
({\rm id} \otimes\Psi_{V,Z})\circ(\Psi_{V,W}\otimes{\rm id}),
\quad \Psi_{V\tens W,Z}=(\Psi_{V,Z}\otimes
{\rm id})\circ({\rm id}\otimes\Psi_{W,Z})
\]
for any three objects. Here we continue to suppress the associativity
morphisms. Note also that we do not assume that
$\Psi_{W,V}\circ\Psi_{V,W}={\rm id}$
for all $V,W$. Naturality means of course that these isomorphisms are
functorial, i.e. all morphisms $f:V\to X$, $g:W\to Y$ in $\calc$
commute
with the braiding in the sense
\[
\Psi_{X,W} \circ (f\otimes {\rm id}_W)
=({\rm id}_W\otimes f)\circ \Psi_{V,W},
\Psi_{V,Y}\circ({\rm id}_V\otimes g)
=(g\otimes{\rm id}_V)\circ\Psi_{V,W}.
\]
A rigid braided monoidal category is a rigid monoidal category which
is
braided. They have been
studied in connection with link invariants in \cite{FreYet:bra}.

Now it is well-known that an invertible Yang-Baxter operator $\calr$
generates
a braided monoidal category. One extends the Yang-Baxter operator from
its
original vector space $V$ to tensor products of $V$ etc in an obvious
way. If
$V$ has an adjunction $\adjoint{}{}{V\sp *}V$ and
$\calr,\mate{\calr}$
are invertible then $\calr$ defines in a similar way a rigid braided
monoidal
category (and leads in fact an Abelian one). This was shown by
Lyubashenko in
his pioneering work~\cite{Lyu:hop}\cite{Lyu:the}, and is by now
well-known. In
fact, we will give for our limited purposes a simplified treatment in
which one
more invertibility is assumed for convenience. Thus we say that a
Yang-Baxter
operator is {\em dualisable} or `closed' if $V$ has an adjunction and
$\calr,\mate{\calr}$ and $\mate{(\calr\sp {-1})}$ are invertible. In
principle,
this last assumption is redundant according to the results effectively
contained in \cite{Lyu:the}.

Given a dualisable Yang-Baxter operator $\calr$ on $V$, the first step
is to
note that it extends to a Yang-Baxter operator $\calr\sp e$ say, on
$V\oplus
V\sp *$. The extension is cf~\cite{Lyu:hop}\cite{Lyu:the}
\begin{equation}
\label{extension1}
\calr\sp e|_{V\ot V} := \calr,\
\calr\sp e|_{V\ot V\sp *} := {\mate \calr}\sp {-1},\
\calr\sp e|_{V\sp *\ot V} := \mate{{\calr\sp {-1}}}\
,\quad
\calr\sp e|_{V\sp *\ot V\sp *} := \calr\sp *.
\end{equation}
That this obeys the Yang-Baxter equations follows from applying the
adjunctions
above to the Yang-Baxter equations for $\calr$.

In the following theorem we suppose
$\adjoint{}{}{X\sp *}X$ and
$\adjoint{}{}{Y\sp *}Y$. Let $(X\oplus Y)\sp *
:= X\sp *\oplus Y\sp *$ and define
$\coev :\bk \to (X\oplus Y)\ot(X\oplus Y)\sp *$ by $\coev(1):=
\coev_X(1) \oplus \coev_Y(1)$ and let $\ev: (X\oplus
Y)\sp *\ot (X\oplus Y)\to \bk$ be defined by
\[
\ev|_{X\sp *\ot X}:=\ev_X,\
\ev|_{Y\sp *\ot Y}:=\ev_Y\
,\quad
\ev|_{(X\sp *\ot Y)\oplus (Y\sp *\ot X)} := 0.
\]
Then $\adjoint{}{}{(X\oplus Y)\sp *}{(X\oplus Y)}$ by these maps
and we have the following theorem.
\begin{theorem}
\label{dualisability_of_glueing}
Let $(X,\calr)$ and $(Y,\calr')$ be elements of $\hq$. Then
the Yang-Baxter operator $\calr\oplusq\calr'$ is dualisable if and
only
if both $\calr$ and $\calr'$ are.
\end{theorem}

\noindent
{\bf Proof.}
The proof is very easy. First, let us agree that the
symbols $x$, $\ox$, $y$ and $\oy$ will denote arbitrary
elements of $X$, $X\sp *$, $Y$ and $Y\sp *$ respectively and let
$\calq =\calr\oplus_q\calr'$. Then we
have the following formulas which can be easily obtained
immediately from the definitions:
\begin{eqnarray*}
\mate{\calq }(\ox\ot x)&=&\mate{\calr}(\ox\ot x),\\
\mate{\calq }(\oy\ot y)&=&\mate{\calr'}(\oy\ot y)
+\aaa\cdot \ev_Y(\oy\ot y) \cdot\coev_X(1),\\
\mate{\calq }(\ox\ot y)&=&y \ot\ox\ ,\quad
\mate{\calq }(\oy\ot x)=x \ot \oy,
\end{eqnarray*}
and
\begin{eqnarray*}
\mate{{\calq \sp {-1}}}(\ox\ot x)&=&\mate{{\calr\sp {-1}}}(\ox\ot x)
-\aaa\cdot\ev(\ox \ot x)\cdot\coev_Y(1),\\
\mate{{\calq \sp {-1}}}(\oy\ot y)&=&\mate{{\calr'\sp {-1}}}(\oy\ot
y),\\
\mate{{\calq \sp {-1}}}(\ox\ot y)&=&y\ot \ox\ ,\quad
\mate{{\calq \sp {-1}}}(\oy\ot x)=x\ot \oy.
\end{eqnarray*}
Now it is immediate to see that the generalized adjoints $\mate{\calq
}$
and $\mate{{\calq \sp {-1}}}$ are invertible if and only if
 $\mate{\calr}$, $\mate{\calr'}$, $\mate{{\calr\sp {-1}}}$
and $\mate{{\calr'\sp {-1}}}$ are invertible.
\qed

The equation~(\ref{extension1}) gives the following explicit
formulas for the extension of the glueing:
\[
(\calr\oplusq\calr')\sp e|_{(X\oplus Y)\ot (X\oplus
Y)}=(\calr\oplusq\calr') ,\
(\calr\oplusq\calr')\sp e|_{(X\oplus Y)\sp *\ot (X\oplus Y)\sp
*}=(\calr\oplusq\calr')\sp *,
\]
$(\calr\oplusq\calr')\sp e|_{(X\oplus Y)\sp *\ot (X\oplus Y)}$ is
given by
\begin{eqnarray*}
(\calr\oplusq\calr')\sp e(\ox\ot x)&=&\calr\sp e(\ox\ot x)
-\aaa\cdot\ev_X(\ox \ot x)\cdot\coev_Y(1),\\
(\calr\oplusq\calr')\sp e(\oy\ot y)&=&\calr'\sp e(\oy\ot
y),\\
(\calr\oplusq\calr')\sp e(\ox\ot y)&=&y\ot \ox,\
(\calr\oplusq\calr')\sp e(\oy\ot x)=x\ot \oy,
\end{eqnarray*}
and $(\calr\oplusq\calr')\sp e|_{(X\oplus Y)\ot (X\oplus Y)\sp *}$ is
given by
\begin{eqnarray*}
(\calr\oplusq\calr')\sp e(x\ot \ox)&=&\calr\sp e(x\ot \ox),\\
(\calr\oplusq\calr')\sp e(y\ot \oy)&=&\calr'\sp e(y\ot \oy)-
\aaa\cdot\ev_Y(\calr'\sp e(y\ot\oy))\cdot
\calr\sp e(\coev_X(1)),\\
(\calr\oplusq\calr')\sp e(x\ot \oy)&=&\oy\ot x\ ,\quad
(\calr\oplusq\calr')\sp e(y\ot \ox)=\ox\ot y.
\end{eqnarray*}

\begin{example}{\rm\
The \YB\ operator $\rsun 1$ on $\bk\ot\bk$ given by
$\rsun1(1\ot1):=q\cdot (1\ot1)$, $q\not=0$, is dualisable, we
have clearly
\[
\mate{\rsun1}=\mbox{multiplication by $q$,}\quad
\mate{{{\rsun1}\sp {-1}}} =\mbox{multiplication by $q\sp {-1}$.}
\]
As we saw in Example~\ref{su(n)}, the
classical \YB\ operator $\rsun n$ is, for $n\geq 1$, build
inductively from $\rsun1$ and
Theorem~\ref{dualisability_of_glueing} says that this operator
is dualisable. It follows
from the same kind of arguments that also the non-standard
\YB\ operator $\calr_{{gl}(1|1)}$ from Example~\ref{AlexConway}
is dualisable.
}\end{example}

Now we come to more categorical considerations. Thus suppose that we are
given a
Yang-Baxter operator $\calr$ on $V$ such that it extends to a rigid
braided
monoidal category containing $V$, with $\Psi_{V,V}=\calr$.
It us immediate that $\calr$ is dualisable and
$\Psi_{V,V\sp *},\Psi_{V\sp *,V\sp *},\Psi_{V\sp *,V}$ are provided by
$\calr{}\sp e$ above.
Some other immediate observations can be obtained at once in the form
of the
following elementary propositions.

\begin{proposition}
\label{u_V}
Suppose that $\calr$ extends to the structure of a braided
rigid category with $\Psi_{V,V}=\calr$ and let
$\adjoint{}{}{V\sp *}V$ and
$\adjoint{}{}{V\sp {**}}{V\sp *}$ be the
adjunctions with their associated evaluation and coevaluation maps. Put
\begin{eqnarray*}
u_V&:=&
(\ev_V\ot{\rm id})(\Psi_{V,V\sp *}\ot{\rm id})({\rm id}\ot\coev_{V\sp *})
:V\to V\sp {**}
\\
v_V&:=&
(\ev_{V\sp *}\ot{\rm id})({\rm id}\ot\Psi_{V, V\sp *})({\rm id}\ot\coev_V)
:V\sp {**}\to V.
\end{eqnarray*}
Then $u_V$ and $v_V$ are invertible morphisms and their inverses
are given by
\begin{eqnarray*}
u_V\sp {-1}&=&
({\rm id}\ot\ev_{V\sp *})
(\Psi_{V\sp {**},V}\ot{\rm id})({\rm id}\ot\coev_V)
\\
v_V\sp {-1}&=&
(\ev_V\ot{\rm id})({\rm id}\ot\Psi_{V\sp {**},
V})(\coev_{V\sp *}\ot{\rm id}).
\end{eqnarray*}
\end{proposition}

\begin{proposition}
\label{properties_of_braiding}
Let $\calr$ extend to the structure of a rigid braided monoidal category as in
Proposition~\ref{u_V}, and define by induction $\dual V0:=V$ and
$\dual V{i+1}:=({\dual Vi})\sp *$ for $i\geq 0$. The braiding has the following
properties.
\begin{enumerate}
\item
$\Psi_{V,V}$ is a dualisable \YB\ operator,
\item
$\Psi_{\dual V{i+1},\dual V{j+1}} =
\left(\Psi_{\dual V{i},\dual V{j}}\right)\sp *,\
i,j\geq 0$,
\item
$\Psi_{\dual V{i+2}, V}=({\rm id}\ot v\sp {-1}_{\dual
Vi})\Psi_{\dual Vi, V}(v_{\dual Vi}\ot{\rm id})$,
for $i\geq 0$,
\item
$\Psi_{V,\dual V{i+2}}=(v\sp {-1}_{\dual Vi}\ot{\rm id})\Psi_{V,\dual
Vi}({\rm id}\ot v_{\dual Vi})$,
for $i\geq 0$.
\end{enumerate}
The last two equations hold also with $u_V$ in place of $v_V$.
\end{proposition}

Notice that equations 3.~and 4.~of
this proposition say exactly that
the maps $v_{\dual Vi}$ commute with the braiding, as
they must
since they are morphisms. That they are morphisms is evident since, by
assumption,
$\ev,\coev$ are morphisms. Notice also that
Proposition~\ref{properties_of_braiding}
together with formula~(\ref{extension1}) enables one to express
inductively $\Psi_{\dual Vi, \dual Vj}$, for any
$i,j\geq 0$, via $\Psi_{V, V}$. Thus we arrive at the
following proposition.

\begin{proposition} (cf~\cite{Lyu:hop}\cite{Lyu:the})
Let $V$ be a vector space with adjunction $V\sp *$ and denote by $\calc(V,V\sp
*)$ the full
rigid subcategory of $\Vect$ generated by $V,V\sp *$.
Let $(V,\calr)$ be a \YB\ operator. Then $\calr$ induces on
$\calc(V,V\sp *)$ the structure of a braided rigid category if and
only if $\calr$ is dualisable.
\end{proposition}

\noindent
{\bf Proof.}
We have already explained one direction: suppose that $\calr$ extends
to a
braiding $\Psi$ to give the structure of a
rigid
braided monoidal category. Especially, it extends to a braiding
between $V$ and
$V\sp*$ which commutes with the adjunctions. One deduces easily that
$\calr$ is
dualisable.

On the other hand, suppose that $\calr$ is dualisable. By definition,
we already have the braiding between $V$ and $V\sp*$ having the
desired
properties. Next, one can write out the braid relations or mixed
quantum-Yang-Baxter equations on the space $V\tens V\sp *\tens V$ and
applying
$\ev_V,\coev_V$ one can deduce that $v\sp {-1}$ exists. We can then define
the braiding between $V$ and
$V\sp{**}$ by identifying $V$ and $V\sp{**}$ via this map. In a
similar way one
extends inductively the braiding to `higher'
duals using the `${\bbb Z}_2$-periodicity'
$\dual V{i+2}\cong \dual V{i}$ induced by $v_{\dual Vi}$.
Formally, this
means extending $\calr$ using the formulas of
Proposition~\ref{properties_of_braiding}. \qed

Summing up the above remarks, we see that we can take the
existence of a braided rigid category generated by $V$ as an
intrinsic definition of the dualisability. With some more
work~\cite{Lyu:the}
one obtains from this an Abelian category, and can also improve on the
requirements for dualisability.

Another important construction in this context is that of quantum or
categorical dimension. Thus in any rigid braided monoidal category
there is an
intrinsic categorical trace of any morphism $f:V\to V$
cf~\cite{Sav:cat}
\[ {\rm Trace}_{{\calr}}(f):= \ev_V\circ\Psi_{V\sp
*,V}\circ({\rm id}\tens
f)\circ \coev_V.\]
The {\em categorical dimension} of any object is defined as the trace of the
identity
morphism. The use of this categorical construction to understand the so-called
quantum
dimensions in the theory of quantum groups was introduced
in~\cite{Ma:qua} and elsewhere. In our present setting the
objects are
vector spaces and the braiding is given by a dualisable \YB\ operator
$\calr$,
in which case the categorical dimension can also be thought of as a
kind of
trace of $\calr$,
\[
\dim_{\calr}(V):=\trace\calr :=\ev\circ\calr\sp e\circ \coev.
\]

\begin{proposition}
\label{trace}
Let $(X,\calr)$ and $(Y,\calr')$ be two dualisable $q$-Hecke \YB\
operators. Then the maps $u_{X\oplus Y}$ and $v_{X\oplus Y}$
related with the glueing $\calr\oplusq\calr'$ satisfy
\begin{eqnarray*}
&u_{X\oplus Y}(x) = u_X(x),
\quad u_{X\oplus Y}(y)=(1-\aaa\cdot\trace\calr)\cdot
u_Y(y),&
\\
& v_{X\oplus Y}(x)=(1-\aaa\cdot\trace{\calr'})\cdot
v_X(x)\ ,\quad v_{X\oplus Y}(y)=v_Y(y),&
\end{eqnarray*}
for $x\in X$, $y\in Y$. Moreover, the braided trace of the
glueing can be computed as
\[
\trace{\calr\oplusq\calr'}=\trace{\calr}+\trace{\calr'}-\aaa
\cdot\trace{\calr}\cdot\trace{\calr'}.
\]
\end{proposition}

{\bf Proof.}
The proof is based on a direct verification.
For example, we have
\begin{eqnarray*}
u_{X\oplus Y}(x) &=&
(\ev_{X\oplus Y}\otimes {\rm id})
(\Psi_{(X\oplus Y),(X\oplus Y)\sp*}\otimes {\rm id})
({\rm id}\otimes\coev_{(X\oplus Y)\sp*})(x)
\\
&=&(\ev_{X\oplus Y}\otimes {\rm id})
(\Psi_{(X\oplus Y),(X\oplus Y)\sp*}\otimes {\rm id})
(x\otimes (\coev_{X\sp*}\oplus \coev_{Y\sp*}))
\\
&=&(\ev_{X\oplus Y}\otimes {\rm id})
(\Psi_{X,X\sp*}\otimes {\rm id})
(x\otimes \coev_{X\sp*})+
(\ev_{X\oplus Y}\otimes {\rm id})
(\Psi_{X,Y\sp*}\otimes {\rm id})
(x\otimes \coev_{Y\sp*})\\
&=& u_X(x).
\end{eqnarray*}
Here we used the fact that
$\Psi_{(X\oplus Y),(X\oplus Y)\sp*}|_{X\otimes X\sp*}=
\Psi_{X,X\sp*}$ and the fact that $\Psi_{(X\oplus Y),(X\oplus Y)\sp*}$
acts on
$X\otimes Y\sp*$ by interchanging the arguments, which implies that
$(\ev_{X\oplus Y}\otimes {\rm id})
(\Psi_{X,Y\sp*}\otimes {\rm id})
(x\otimes \coev_{Y\sp*})=0$.
The verification of the remaining equations of the proposition is
similar.
\qed

\begin{example}{\rm
\label{trace-of-su(n)}
Let $\rsun n$ be the classical \YB\ operator introduced in
Example~\ref{su(n)}. Clearly $\trace{\rsun1}=q\sp {-1}$ and the
proposition above gives by induction the formula
\[
\trace{\rsun n}=\sum_{i=1}\sp n q\sp {-2i+1}={1-q\sp {-2n}\over 1-q\sp
{-2}}q\sp {-1}.
\]
A similar computation shows that $\trace{\calr_{{gl}(1|1)}}$
of the non-standard \YB\ operator of Example~\ref{AlexConway} is zero.
These
are essentially the
standard results of the quantum dimension of the fundamental
representation of
the associated quantum groups.
}\end{example}

\begin{example}{\rm\
For $V=\mbox{Span}(x_1,\ldots,x_n)$ let
$(x\sp {**}_1,\ldots,x_n\sp {**})$ be the basis of $V\sp {**}$ given
by $x_i\sp {**}(\varphi)=\varphi(x_i)$, $\varphi\in V\sp *$.
Using inductively Proposition~\ref{trace} and the formula
for $\trace{\rsun n}$ computed in the previous example, we
get the following formulas for the maps $u_V$ and $v_V$
related with $\rsun n$:
\[
u_V(x_i) = q\sp {-2i+1}\cdot x\sp {**}_i\
,\quad
v_V(x\sp {**}_i)=q\sp {-2n+2i-1}\cdot x_i,
\]
for $1\leq i\leq n$.
}\end{example}

We close this section by discussing the following notion
related with invariants of oriented framed isotopy classes of
links. Let $\calr$ be a dualisable \YB\ operator on $V$ and
denote, as usual, be $\calr\sp e$ its extension on $V\oplus
V\sp *$. By the {\em double twist} we mean the map
$\chi_{\calr}:V\to V$
defined by
\[
\chi_{\calr}:=({\rm id}\ot\ev)
(\calr\sp e\ot{\rm id})(\calr\sp e\ot{\rm id})(\coev\ot{\rm id})=
({\rm id}\ot\ev)
(\mate{{\calr\sp {-1}}}\ot{\rm id})({\mate{\calr}}\sp {-1}
\ot{\rm id})(\coev\ot{\rm id}).
\]
One says that
$\calr$ is {\em tortile}~\cite{JoyStr:tor} or {\em
ribbon}~\cite{ResTur:rib} if
there exists a map $\theta :V
\to V$ (a {\em ribbon element}) such that $\chi_{\calr}=\theta\sp 2$.
The following proposition shows that the structure of the
double twist of $q$-Hecke \YB\ operators is very simple.

\begin{proposition}
\label{easy_tortile}
Let $\calr$ be a dualisable $q$-Hecke \YB\ operator. Then
\[
\chi_{\calr}= (1-\aaa\trace{\calr})\cdot\id.
\]
\end{proposition}

\noindent
{\bf Proof.}
The $q$-Hecke property can be rewritten as
$\calr\sp {-1}=\calr - \aaa\cdot(\id\ot\id)$. We have from
this $\mate{{\calr\sp {-1}}}\circ {\mate{\calr}}\sp {-1} =
(\id\ot\id)-
\aaa\cdot\mate{(\id\ot\id)}\circ(\mate{\calr})\sp {-1}$. Plugging
it into the formula defining $\chi_{\calr}$ we get the
requisite result.
\qed

Using Proposition~5.6 we easily get the
following corollary.

\begin{corollary}
\label{corollary}
Let $(X,\calr)$ and $(Y,\calr')$ be two $q$-Hecke
dualisable \YB\ operators. Then
\[
\chi_{\glueing}=\left[(1-\aaa\cdot\trace\calr)\cdot
(1-\aaa\cdot\trace{\calr'})\right]\cdot\id.
\]
Thus, the scalar multiples by which the operators $\chi_{\calr}$ act behave
multiplicatively under $\oplus_q$.
\end{corollary}

As an immediate consequence of
Proposition~\ref{easy_tortile} we get that a dualisable
$q$-Hecke \YB\ operator $\calr$ has a twist if and only if the
scalar $(1-\aaa\cdot\trace\calr)$ has a square root. Let
$\lambda_\calr$ be such a square root. Then the twist $\theta$
is necessarily of the form $\theta = \lambda_\calr \cdot
E$, where $E$ is an endomorphism such that $E\sp 2=\id$.
Combining the results above, we obtain the following
proposition.

\begin{proposition}
Let $(\calr,\theta)$ and $(\calr',\theta')$ be two tortile
$q$-Hecke \YB\ operators. Suppose $\theta = \lambda\cdot
E$, $\theta' = \lambda'\cdot E'$, for some endomorphisms
$E$ and $E'$ with $E\sp 2=\id$ and $E'\sp 2=\id$. If we put
$\overline\theta := \lambda\cdot\lambda'\cdot (E\oplus
E')$, then $(\glueing,\overline\theta)$ is also a tortile
\YB\ operator.
\end{proposition}

\begin{example}{\rm\ From Example~\ref{trace-of-su(n)} we get easily
\[
\chi_{\rsun n}= q\sp {-2n}\cdot\id\
,\quad
\chi_{\calr_{{gl}(1|1)}}=\id
\]
(the formula for $\chi_{\rsun n}$ can be also obtained
inductively from Corollary~\ref{corollary}). Both operators
are tortile and we usually take $\theta =q\sp {-n}\cdot \id$
for the first one and $\theta=\id$ for the second one.
These are essentially the standard results for these quantum groups,
cf~\cite{JoyStr:tor}\cite{ResTur:rib}.}\end{example}

\section{Glueing of link invariants}
\setlength{\unitlength}{.9cm}
\thicklines

\newsavebox{\erplus}
\savebox{\erplus}(0,0)[lb]{
\put(0,0){\vector(1,-1){1}}
\put(1,0){\line(-1,-1){.4}}
\put(.4,-.6){\vector(-1,-1){.4}}
}

\newsavebox{\ernula}
\savebox{\ernula}(0,0)[lb]{
\put(0,0){\vector(1,-1){1}}
\put(1,0){\vector(-1,-1){1}}
\put(.5,-.49){\makebox(0,0){$\bullet$}}
}

\newsavebox{\erminus}
\savebox{\erminus}(0,0)[lb]{
\put(1,0){\vector(-1,-1){1}}
\put(0,0){\line(1,-1){.4}}
\put(.6,-.6){\vector(1,-1){.4}}
}

\def\er#1#2#3#4#5{
\begin{picture}(1.8,1)(-.4,-.6)
\put(0,-1.5){\makebox(0,0){$#1$}}
\put(1,-1.5){\makebox(0,0){$#2$}}
\put(0,.5){\makebox(0,0){$#3$}}
\put(1,.5){\makebox(0,0){$#4$}}
\if#5+
\def\xxx{\erplus}
\else
\def\xxx{\erminus}
\fi
\if#50
\def\xxx{\ernula}
\else\fi
\put(0,0){\usebox{\xxx}}
\end{picture}
}

\def\identity#1#2{
\begin{picture}(2,1.5)(-.5,-.6)
\bezier{100}(0,0)(.5,-.5)(0,-1)
\bezier{100}(1,0)(.5,-.5)(1,-1)
\put(-.3,-.5){\makebox(0,0){$#1$}}
\put(1.3,-.5){\makebox(0,0){$#2$}}
\put(1,-1){\vector(1,-1){0}}
\put(0,-1){\vector(-1,-1){0}}
\end{picture}
}

\newsavebox{\UU}
\savebox{\UU}(0,0)[lb]{
\bezier{50}(-0.03,0)(-0.03,-.5)(.47,-.5)
\bezier{50}(.47,-.5)(0.97,-.5)(0.97,0)
}

\newsavebox{\VV}
\savebox{\VV}(0,0)[lb]{
\bezier{50}(-.03,0)(-.03,.5)(.47,.5)
\bezier{50}(.47,.5)(0.97,.5)(0.97,0)
}

\def\THETA{
\begin{picture}(1.5,2)(0,-.1)
\bezier{50}(0.03,0)(0.03,-.5)(.53,-.5)
\bezier{50}(0.03,0)(0.03,.5)(.53,.5)
\bezier{50}(.53,-.5)(1.03,-.5)(1.03,0)
\bezier{25}(.53,.5)(.73,.5)(.93,.27)
\put(1.05,-.4){\vector(0,-1){.6}}
\put(1.045,0){\line(0,1){1}}
\end{picture}
}

\def\THETAINV{
\begin{picture}(1.5,2)(0,-.1)
\bezier{50}(0.03,0)(0.03,-.5)(.53,-.5)
\bezier{50}(0.03,0)(0.03,.5)(.53,.5)
\bezier{50}(.53,.5)(1.03,.5)(1.03,0)
\bezier{25}(.53,-.5)(.73,-.5)(.93,-.27)
\put(1.05,.4){\line(0,1){.6}}
\put(1.045,0){\vector(0,-1){1}}
\end{picture}
}

\def\uparrow{
\begin{picture}(0,2)(0,-.5)
\put(0,-1){\vector(0,1){2}}
\end{picture}
}

\def\ID{
\hskip2mm
\begin{picture}(0,.85)(0,-.3)
\put(0,.25){\vector(0,-1){.85}}
\end{picture}
\hskip2mm
}

\def\EPSILONBAR{
\begin{picture}(1,1)
\put(0,0){\usebox{\UU}}
\put(0,0){\makebox(0,0)[b]{\vector(0,1){.5}}}
\put(1,0){\makebox(0,0)[b]{\line(0,1){.5}}}
\end{picture}
}

\def\EPSILON{
\begin{picture}(1,1)
\put(0,0){\usebox{\UU}}
\put(1,0){\makebox(0,0)[b]{\vector(0,1){.5}}}
\put(0,0){\makebox(0,0)[b]{\line(0,1){.5}}}
\end{picture}
}

\def\ETA{
\begin{picture}(1,1)
\put(0,0){\usebox{\VV}}
\put(0,0){\vector(0,-1){.5}}
\put(1,0){\line(0,-1){.5}}
\end{picture}
}

\def\ETABAR{
\begin{picture}(1,1)
\put(0,0){\usebox{\VV}}
\put(1,0){\vector(0,-1){.5}}
\put(0,0){\line(0,-1){.5}}
\end{picture}
}

\def\GAMMA{
\begin{picture}(1,1)(0,-.5)
\put(.75,.25){\line(1,0){.5}}
\put(.75,-.25){\line(1,0){.5}}
\put(.75,-.25){\line(0,1){.5}}
\put(1.25,-.25){\line(0,1){.5}}
\put(1,0){\makebox(0,0){n}}
\put(.99,.25){\line(0,1){.25}}
\put(.99,-.25){\line(0,-1){.25}}
\put(-.02,.5){\vector(0,-1){1}}
\put(0,.5){\usebox{\VV}}
\put(0,-.5){\usebox{\UU}}
\end{picture}
}

\def\krabicka{
\begin{picture}(0,1)(0,-.2)
\put(-.25,.25){\line(1,0){.5}}
\put(-.25,-.25){\line(1,0){.5}}
\put(-.25,-.25){\line(0,1){.5}}
\put(.25,-.25){\line(0,1){.5}}
\put(0,.25){\line(0,1){.2}}
\put(0,-.25){\line(0,-1){.2}}
\put(0,0){\makebox(0,0){n}}
\end{picture}
}
As we have already mentioned, there is an intimate
relationship between link invariants and solutions of the
\YB\ equation. This suggest the possibility of the
existence of an analog of glueing also for link invariants.
The aim of this last section is to describe such a
construction, based on the state-model approach of
L.~Kauffman~\cite{Kau:phy}.

For an (oriented) link, choose a plane projection without
triple points and with only finitely many double points.
We call such a projection the {\em shadow\/}
of $L$. The isotopy type of $L$ is then determined by its
shadow plus the over/under information at the vertices (the double
points)
of the shadow. By a {\em state\/} of $L$ we mean an
assignment of an element of the set $\{x,y\}$ to each edge
of the shadow. We will consider only states satisfying the
{\em spin-conservation rule\/} meaning that at each double
point of the shadow $S$,
\cskip
\hfill\er jlik0 \quad $i,j,k,l\in \{x,y\}$,\hfill
\endcskip
the equation $s(i)+s(k)=s(j)+s(l)$, with $s$ defined by
$s(x)=-1/2$, $s(y)=+1/2$, is satisfied.

Let $L$ be a link and $\sigma$ a state as above.
It is convenient to classify the double points of the shadow and the
related over/under information in a given state into the following types:
\cskip
Type~I:\hfill
$\er xxxx+ \mbox{ and } \er yyyy+ ,$\hfill\hphantom{Type~I:}
\endcskip
\vskip-10pt
\cskip
Type~II:\hfill
$\er xxxx- \mbox{ and } \er yyyy- $,\hfill\hphantom{Type~II:}
\endcskip
\vskip-10pt
\cskip
Type III:\hfill
$\er xyxy+ \mbox{ and } \er yxyx-, $\hfill\hphantom{Type III:}
\endcskip
\vskip-10pt
\cskip
Type IV:\hfill
$\er yxyx+,$\hfill\hphantom{Type IV:}
\endcskip
\vskip-10pt
\cskip
Type V:\hfill
$\er xyxy- ,$\hphantom{Type V:}\hfill
\endcskip
\vskip-10pt
\cskip
Type VI:\hfill$
\er xyyx+,\quad
\er yxxy+,\quad
\er xyyx-\mbox{ and }
\er yxxy-.
$\hphantom{Type VI:}\hfill
\endcskip
\vskip5mm

By our spin-conservation rule, listed above are all possible
situations.

Let $L\sp \sigma$ be the singular (= possibly with double
points) oriented link and state on it constructed from $L$ and
$\sigma$ by the
replacements:

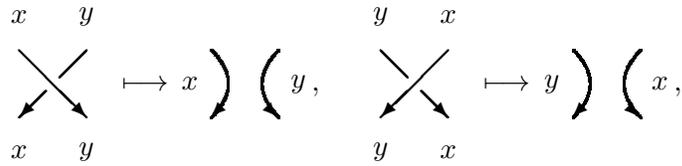
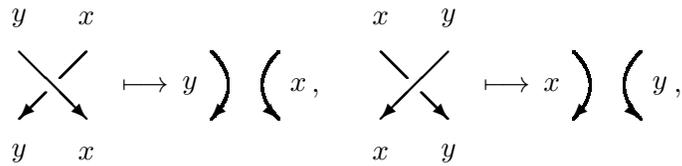
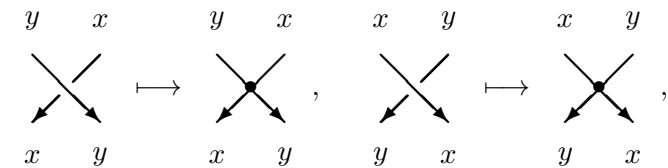
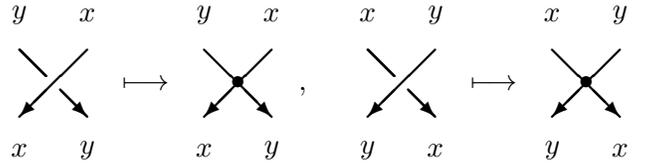
\cskip
\hfill
$\er xyxy+ \lmt \identity xy, \quad \er yxyx- \lmt\identity
yx,$\hfill
\endcskip
\vskip-10pt
\cskip
\hfill$
\er yxyx+\lmt \identity yx,\quad
\er xyxy- \lmt \identity xy,
$\hfill
\endcskip
\vskip-10pt
\cskip
\hfill$
\er xyyx+\lmt \er xyyx0,\quad
\er yxxy+\lmt \er yxxy0,
$\hfill
\endcskip
\vskip-10pt
\cskip
\hfill$
\er xyyx- \lmt \er xyyx0,\quad
\er yxxy-\lmt \er yxxy0,
$\hfill
\endcskip
\vskip5mm
the other types of vertices remaining unchanged.

Let $L\sp \sigma_y$ be the component of $L\sp \sigma$ colored by
$x$ and let $L\sp \sigma_y$ have the similar meaning for coloring by $y$.
Notice
that the intersection $L\sp \sigma_x\cap L\sp \sigma_y$ need not
be void, but $L\sp \sigma_x$ and $L\sp \sigma_y$ are honest
oriented links (with no self-intersections).

Denote by $V$ the set of vertices of the
shadow $S$. Using our above classification of the types of vertices relative to
our state, we define the {\em weight\/}
$W\sp{\sigma}(v)$ of $v\in V$ by
\[
W\sp \sigma(v)=\left\{
\begin{array}{ll}
1,& \mbox{for types I, II and VI,}\\
0,& \mbox{for type III,}\\
\aaa,& \mbox{for type IV,}\\
-\aaa,& \mbox{for type V.}
\end{array}
\right.
\]

Let $L$ be an oriented link. Making the replacements
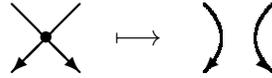
\cskip\vskip-5mm
\hfill
$\er{}{}{}{}0\lmt \identity{}{}$
\hfill
\endcskip

\vskip-3mm
in the shadow $S$ of $L$, we get a finite set of oriented
circuits in the plane. Define $n(L):=$ number of anticlockwise oriented
circuits $-$ number of clockwise oriented circuits.

Consider a $q$-Hecke tortile \YB\ operator
$(\calr,\theta)$.
It is well-known (see, for
example,~\cite{JoyStr:tor})
that $(\calr,\theta)$ gives rise to a
framed oriented link isotopy invariant, say $\kappa(-)$.
The idea of the construction of $\kappa$ is to
interpret a given link as a map in a certain free tortile category,
the interpretation being based on the assignments
\par\vskip0mm
\[
\er{}{}{}{}+\lmt\ \calr,\
\er{}{}{}{}-\lmt\ \calr\sp {-1},\hskip4mm
\
\ETA\ \ \lmt\ \ \coev,
\]
\par
\[
\EPSILONBAR\ \ \lmt\ \ \ev,\hskip4mm
\ETABAR\ \ \lmt\ \ (\id\ot\theta\sp {-1})\circ\calr\sp
e\circ\coev,\hskip4mm
\EPSILON\ \ \lmt\ \ \ev\circ\calr\sp e\circ(\id\otimes\theta\sp
{-1}).
\]
\par
Denote
\[
\vartheta:=\THETA,\quad \vartheta\sp{-1}:=\THETAINV.
\]
\par\vskip2mm
and let \hskip3mm $\krabicka$ \hskip3mm
be the $n$-fold composition of $\vartheta$ for
$n\geq 0$ and the $(-n)$-fold composition of $\vartheta\sp{-1}$ for
$n<0$. Let, finally, $\Gamma_n$ be the link
\par\vskip3mm
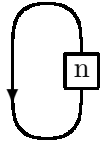
\[
\GAMMA
\]
\par
We can easily prove using the results of Section~5, the following
proposition.

\begin{proposition}
\label{invariant_of_gamma}
Let $(\calr,\theta)$ and $(\calr',\theta')$ be two $q$-Hecke tortile
\YB\ operators and suppose that $\theta = \lambda\cdot\id$ and
$\theta' = \lambda'\cdot\id$ for some scalars $\lambda$ and
$\lambda'$. Let $\kappa$, $\kappa'$ and $\omega$
be the oriented framed link isotopy invariants
related with the operators
$\calr$, $\calr'$ and $\calr\oplusq\calr'$, respectively.
Then
\[
\omega(\Gamma_n)=
\lambda'\sp{n-1}\kappa(\Gamma_n)+\lambda\sp{n+1}\kappa'(\Gamma_n).
\]
\end{proposition}

{\bf Proof.} From the assumption $\theta = \lambda\cdot\id$ we
easily have that $\kappa(\Gamma_n)= \lambda\sp n\kappa(\Gamma_0)$
(notice that $\Gamma_0$ is
the oriented circle). Immediately from the definition we get that
$\kappa(\Gamma_0)=
\ev\circ\calr\sp e\circ(\id\otimes\theta\sp{-1})\circ \coev =
\lambda\sp{-1}\cdot\trace{\calr}$. This, and a similar computation gives
\[
\kappa(\Gamma_n)=\lambda\sp{n-1}\cdot\trace{\calr},\qquad
\kappa(\Gamma'_n)=\lambda'\sp{n-1}\cdot\trace{\calr'}.
\]
Recall that we consider the glueing $\calr\oplusq\calr'$ as a tortile
Yang-Baxter operator with the twist given by $\overline\theta =
\lambda\lambda'\cdot\id$ The similar considerations as above
together with Proposition \ref{trace} give that
\begin{eqnarray*}
\omega(\Gamma_n)&=&
\lambda\sp{n-1}\lambda'\sp{n-1}\cdot\trace{\calr\oplusq\calr'}
\\
&=&
\lambda\sp{n-1}\lambda'\sp{n-1}\cdot
[\trace\calr +\trace{\calr'}-\aaa\cdot
\trace{\calr}\trace{\calr'}]=
\\
&=&\lambda'\sp{n-1}[\lambda\sp{n-1}\trace\calr]+
\lambda\sp{n-1}[\lambda'\sp{n-1}\trace\calr'][1-\aaa\cdot\trace\calr]
\end{eqnarray*}
which, together with the formula $\lambda\sp2=1-\aaa\cdot\trace\calr$
(see the comments following Corollary \ref{corollary}) gives the
requisite equation.
\qed

Let $(\calr,\theta)$ and $(\calr',\theta')$ be as in
Proposition~\ref{invariant_of_gamma} and let $\kappa$ and $\kappa'$
be the related invariants.

\begin{theorem} \label{glueing-of-invariants}
Let $\kappa,\kappa'$ be invariants of framed oriented isotopy
and extend the definition by
$\kappa(\emptyset)=\kappa'(\emptyset):=1$ for the empty link
$\emptyset$.
Suppose, moreover, that
\[
\kappa(\vartheta)=\lambda\cdot\kappa(\ID),\
\kappa(\vartheta\sp{-1})=\lambda\sp{-1}\cdot\kappa(\ID),\
\kappa'(\vartheta)=\lambda'\cdot\kappa'(\ID),\
\kappa'(\vartheta\sp{-1})=\lambda'\sp{-1}\cdot\kappa'(\ID),
\]
for $\lambda,\lambda'\in \bk$. Then their {\em glueing\/}
$\kappa\oplusq\kappa'$ defined by
\[
(\kappa\oplusq\kappa')(L):=
\sum_{\scriptsize\mbox{states }\sigma}
\kappa(L\sp \sigma_x)\kappa'(L\sp
\sigma_y)
(\prod_{v\in V}W\sp \sigma(v))\lambda'\sp {-n(L\sp \sigma_x)}
\lambda\sp {n(L\sp \sigma_y)}
\]
is an invariant of framed oriented isotopy. Moreover,
if $\kappa,\kappa'$ correspond to $q$-Hecke Yang-Baxter operators as
in
Proposition~6.1 then their glueing $\kappa\oplusq\kappa'$ corresponds
to
$\calr\oplusq\calr'$.
\end{theorem}
\noindent
{\bf Proof.}
We shall prove first that the stated formula for $\kappa\oplus_q\kappa'$ really
defines a framed oriented isotopy invariant, which is the
same as to prove that it is invariant
under oriented framed Reidemeister
moves~\cite[page~330]{Yet:fra}.
This can be done by a direct verification, using the
invariance of $\kappa$ and $\kappa'$ and the assumed values on
$\vartheta,\vartheta\sp {-1}$. These values are the ones for the
invariants
corresponding to the Yang-Baxter operators in Proposition~6.1 where
$\theta =
\lambda\cdot\id$ and
$\theta'=\lambda'\cdot\id$.
It remains to prove that
$\kappa\oplusq\kappa'$ is related with the glueing
$\calr\oplusq\calr'$, in other words, we shall prove that $\omega=
\kappa\oplusq\kappa'$, where $\omega$ has the same meaning as in
Proposition~\ref{invariant_of_gamma}.

Observe that both oriented framed link isotopy invariants $\omega$
and $\kappa\oplusq\kappa'$
satisfy the `skein relations' in the form
of~\cite[page~23]{Ma:qua} -- for
$\omega$ it follows from the fact that $\calr\oplusq\calr'$ is
$q$-Hecke while for $\kappa\oplusq\kappa'$ it follows from the
definition and from the fact that $\kappa$ and
$\kappa'$ share this property, since $\calr$ and $\calr'$ are
$q$-Hecke.
Invoking the `skein template algorithm'
of~\cite[pages 57--63]{Kau:phy} we see that it is
enough to verify that
$\omega(\Gamma_n)=(\kappa\oplusq\kappa')(\Gamma_n)$ for all $n$, but
the left hand side was already computed in of Proposition~6.1 and it
is easy to
see that it coincides with the value of $\kappa\oplus_q\kappa'$ from
its
definition.
\qed

\begin{example}{\rm\
\label{writhe}
Let us recall first the definition of the {\em writhe\/}
$w(L)$ of an oriented link $L$. We put
\[
w(L)=\sum_{v\in V}\epsilon(v),
\]
where the summation is taken over the set of vertices of
the shadow of $L$ and $\epsilon$ is defined by

\vskip-10mm
\cskip
\hfill
$\epsilon\left(\hskip-2mm\rule{0cm}{.65cm}\right.
\er{}{}{}{}+\left.\rule{0cm}{.65cm}\hskip-2mm\right):=+1\
,\quad
\epsilon\left(\hskip-2mm\rule{0cm}{.65cm}\right.
\er{}{}{}{}-\left.\rule{0cm}{.65cm}\hskip-2mm\right):=-1.$
\hfill
\endcskip

\vskip-5mm
Consider now $\rsun 1$ as a tortile \YB\ operator with
$\theta=q\cdot \id$. It is easy to show that the related invariant
is given by $\kappa(L)=q\sp {w(L)}$. We show that the
formula in Theorem~\ref{glueing-of-invariants}
for the glueing $\kappa\oplusq\kappa$ gives the usual state
model for the $\rsun2$-Jones invariant, as predicted by
Theorem~\ref{glueing-of-invariants}. We have
\begin{equation}
\label{kappa}
(\kappa\oplusq\kappa)(L):=
\sum_\sigma q\sp {w(L\sp \sigma_x)+w(L\sp \sigma_y)}
(\prod_{v\in V}W\sp \sigma(v))q\sp {-n(L\sp \sigma_x)}
q\sp {n(L\sp \sigma_y)}.
\end{equation}
Recall our classification of the types of vertices of the
shadow of $L$ and define, for each vertex $v\in V$,
\[
U\sp \sigma(v):=
\left\{
\begin{array}{ll}
q,& \mbox{for type I,}\\
q\sp {-1},& \mbox{for type II,}\\
1, &\mbox{for types III--VI.}
\end{array}
\right.
\]
It is immediate from the definition that
\[
q\sp {w(L\sp \sigma_x)+w(L\sp \sigma_y)} =\prod_{v\in V}U\sp
\sigma(v).
\]
Looking at $W\sp \sigma_{\rsun 2}(v):= U\sp \sigma (v)\cdot W\sp
\sigma (v)$,
we see
that
$W\sp \sigma _{\rsun 2}(v)$ is the usual Boltzmann weight
given by the formula
\[
W\sp \sigma_{\rsun 2}(v)=
\left\{
\begin{array}{ll}
q, &\mbox{for type I,}\\
q\sp {-1}, &\mbox{for type II,}\\
0, &\mbox{for type III,}\\
\aaa, &\mbox{for type IV,}\\
-\aaa, &\mbox{for type V,}\\
1, &\mbox{for type VI}
\end{array}
\right.
\]
Summing up the above observations we see that the
formula~(\ref{kappa}) is identical with Kauffman's state-model
definition of the $\rsun 2$-Jones invariant. We used
the conventions in~\cite[par.~2.2]{Ma:qua}.
}\end{example}

\end{document}